\documentclass[fleqn,usenatbib]{mnras}

\usepackage{newtxtext,newtxmath}

\usepackage[T1]{fontenc}

\DeclareRobustCommand{\VAN}[3]{#2}
\let\VANthebibliography\thebibliography
\def\thebibliography{\DeclareRobustCommand{\VAN}[3]{##3}\VANthebibliography}

\usepackage{graphicx}	
\usepackage{amsmath}	

\usepackage{bm}

\usepackage{CJKutf8}

\newcommand{\vect}[1]{\bm{#1}}
\newcommand{\matr}[1]{\mathbfss{#1}}
\newcommand{\vth}{v_{\text{th}}}
\newcommand{\St}{\text{St}}
\newcommand{\atantwo}{\text{atan2}}
\DeclareMathOperator{\arcsinh}{arcsinh}

\title[Birdie-like Alignment]{Badminton Birdie-Like Aerodynamic Alignment of Drifting Dust Grains by Subsonic Gaseous Flows in Protoplanetary Disks}

\author[Lin et al.]{
Zhe-Yu Daniel Lin (\begin{CJK*}{UTF8}{bkai}林哲宇\end{CJK*}),$^{1,2}$\thanks{E-mail: zdl3gk@virginia.edu or zlin@carnegiescience.edu}
Zhi-Yun Li,$^{1}$
Haifeng Yang,$^{3}$
Leslie W. Looney,$^{4,5}$
Ian W. Stephens,$^{6}$
\newauthor
Manuel Fern\'andez-L\'opez,$^{7}$
Rachel E. Harrison$^{8}$
\\
$^{1}$Department of Astronomy, University of Virginia, 530 McCormick Rd., Charlottesville 22904, Virginia, USA\\
$^{2}$Earth and Planets Laboratory, Carnegie Science, 5241 Broad Branch Rd. NW, Washington, DC 20015, USA \\
$^{3}$Institute for Astronomy, School of Physics, Zhejiang University, Hangzhou, Zhejiang 310027, China\\
$^{4}$Department of Astronomy, University of Illinois, 1002 W Green St., Urbana, Illinois 61801, USA\\
$^{5}$National Radio Astronomy Observatory, 520 Edgemont Rd., Charlottesville, Virginia 22903, USA\\
$^{6}$Department of Earth, Environment, and Physics, Worcester State University, Worcester, Massachusetts 01602, USA\\
$^{7}$Instituto Argentino de Radioastronom{\'i}a, CCT-La Plata (CONICET), C.C.5, 1894, Villa Elisa, Argentina\\
$^{8}$School of Physics and Astronomy, Monash University, Clayton VIC 3800, Australia\\
}

\date{Accepted XXX. Received YYY; in original form ZZZ}

\pubyear{2015}

\begin{document}
\label{firstpage}
\pagerange{\pageref{firstpage}--\pageref{lastpage}}
\maketitle

\begin{abstract}

Recent (sub)millimeter polarization observations of protoplanetary disks reveal toroidally aligned, effectively prolate dust grains large enough (at least $\sim 100$~$\mu$m) to efficiently scatter millimeter light. 
The alignment mechanism for these grains remains unclear. 
We explore the possibility that gas drag aligns grains through gas-dust relative motion when the grain's center of mass is offset from its geometric center, analogous to a badminton birdie's alignment in flight. 
A simple grain model of two non-identical spheres illustrates how a grain undergoes damped oscillations from flow-induced restoring torques which align its geometric center in the flow direction relative to its center of mass. 
Assuming specular reflection and subsonic flow, we derive an analytical equation of motion for spheroids where the center of mass can be shifted away from the spheroid's geometric center. 
We show that a prolate or an oblate grain can be aligned with the long axis parallel to the gas flow when the center of mass is shifted along that axis. 
Both scenarios can explain the required effectively prolate grains inferred from observations. 
Application to a simple disk model shows that the alignment timescales are shorter than or comparable to the orbital time. 
The grain alignment direction in a disk depends on the disk (sub-)structure and grain Stokes number (St) with azimuthal alignment for large St grains in sub-Keplerian smooth gas disks and for small St grains near the gas pressure extrema, such as rings and gaps.

\end{abstract}

\begin{keywords}
polarization -- protoplanetary disks 
\end{keywords}



\section{Introduction}
Dust grains within gaseous protoplanetary disks are the raw material for forming planets \citep[e.g.][]{Drazkowska2023ASPC..534..717D}. 
Their studies have been revolutionized by the Atacama Large Millimeter/submillimeter Array (ALMA), especially through disk-scale observations of linear continuum polarization \citep[e.g.][]{Kataoka2016ApJ...831L..12K, Stephens2017ApJ...851...55S, Lee2018ApJ...854...56L, Girart2018ApJ...856L..27G, Ohashi2018ApJ...864...81O, Alves2018A&A...616A..56A, Sadavoy2018ApJ...869..115S, Bacciotti2018ApJ...865L..12B, Dent2019MNRAS.482L..29D, Takahashi2019ApJ...872...70T, Harrison2019ApJ...877L...2H, Sadavoy2019ApJS..245....2S, Stephens2020ApJ...901...71S, Aso2021ApJ...920...71A, Tang2023ApJ...947L...5T, Stephens2023Natur.623..705S, Lin2024MNRAS.528..843L, Liu2024ApJ...963..104L, Harrison2024ApJ...967...40H}. The origin of disk polarization is often attributed to scattering when dust grains have grown large enough to efficiently scatter millimeter/sub-millimeter light. Measuring polarization can constrain the properties of dust grains, like the distribution, grain sizes, porosity, etc \citep[e.g.][]{Kataoka2015ApJ...809...78K, Yang2016MNRAS.456.2794Y, Kataoka2016ApJ...820...54K, Yang2017MNRAS.472..373Y, Kirchschlager2020A&A...638A.116K, Lin2020MNRAS.496..169L, Yang2020ApJ...889...15Y, Zhang2023ApJ...953...96Z, Zamponi2024A&A...682A..56Z, Yang2024ApJ...963..134Y}. Alternatively, if grains are aligned, then the thermal emission is intrinsically polarized \citep[e.g.][]{vandeHulst1957lssp.book.....V, Yang2016MNRAS.460.4109Y, Kirchschlager2019MNRAS.488.1211K, Lin2020MNRAS.493.4868L, Lin2022MNRAS.512.3922L}.

Polarized thermal emission from aligned grains has been used to trace the magnetic field in the envelopes around protostars. Polarization successfully measured the expected hour-glass morphology of the magnetic field \citep[e.g.][]{Girart2006Sci...313..812G, Stephens2013ApJ...769L..15S, Maury2018MNRAS.477.2760M, Kwon2019ApJ...879...25K, Huang2024ApJ...963L..31H}. The widely accepted explanation is through radiative alignment torques, i.e., RATs \citep{Dolginov1972Ap&SS..18..337D, Dolginov1976Ap&SS..43..257D, Draine1996ApJ...470..551D, Draine1997ApJ...480..633D, Lazarian2007MNRAS.378..910L} which allow grains to be aligned to the magnetic field or radiation field through internal and external alignment (see, e.g., \citealt{Lazarian2007JQSRT.106..225L}, \citealt{Lazarian2015psps.book...81L} for a review). Internal alignment refers to the alignment of the angular momentum, $\vect{J}$, to the axis of maximum moment of inertia due to energy dissipation \citep{Purcell1979ApJ...231..404P, Lazarian1999MNRAS.303..673L, Hoang2009ApJ...697.1316H, Hoang2022AJ....164..248H}. As a result, the long axis of the grain should be perpendicular to $\vect{J}$ and the polarization of an ensemble of grains should appear effectively oblate. External alignment occurs when $\vect{J}$ becomes aligned with an external field, like the magnetic field $\vect{B}$ or the radiation field \citep{Lazarian2007MNRAS.378..910L, Lazarian2015psps.book...81L}. When producing polarization after internal and external alignments, the grains as an ensemble should be effectively oblate with their short axes aligned along the external field. 

In the disk-scales, however, recent observations have revealed compelling evidence for the presence of grains that appear effectively prolate with their long axes aligned toroidally around the disk. For example, multiwavelength observations of HL~Tau have unveiled transitions in the polarization morphology from the Very Large Array (VLA) Q-band ($\lambda=7.1$~mm) to ALMA Bands 3, 4, 5, 6, and 7 ranging from $\lambda=$3.1~mm to 0.87~mm \citep{Stephens2014Natur.514..597S, Stephens2017ApJ...851...55S, Lin2024MNRAS.528..843L}. The change in pattern cannot be solely attributed to scattering. In particular, Band 3 ($\lambda=$3.1~mm) shows an azimuthally oriented polarization pattern that was shown to require toroidally aligned, effectively prolate grains based on the azimuthal variation in the polarization \citep{Kataoka2017ApJ...844L...5K, Mori2021ApJ...908..153M, Yang2019MNRAS.483.2371Y}. 
Since the optical depth decreases with increasing wavelength, scattering can become less of a factor at longer wavelengths. Multiwavelength models have shown that scattering of toroidally aligned, prolate grains can produce the observed polarization transition \citep{Lin2022MNRAS.512.3922L, Lin2024MNRAS.528..843L}. 
In addition, recent high angular resolution observations ($5$~au resolution) at ALMA Band~7 ($\lambda=0.87$~mm) revealed azimuthally oriented polarization in the first gap, but signatures of scattering polarization in the rings. \cite{Stephens2023Natur.623..705S} demonstrated that by incorporating optical depth changes between rings and gaps, scattering of toroidally aligned, prolate grains can also explain the observed polarization substructure. 

Signatures of toroidally aligned, prolate grains akin to those observed from HL~Tau also exist in disks around other sources. 
Haro~6-13, V892~Tau, DG~Tau, and GG~Tau show polarization that is predominantly parallel to the disk minor axis (a signature of scattering) at the shorter wavelength, $\lambda \sim 0.9$~mm, and azimuthally oriented polarization at the longer wavelength, $\lambda\sim3$~mm
\citep{Bacciotti2018ApJ...865L..12B, Harrison2019ApJ...877L...2H, Tang2023ApJ...947L...5T, Ohashi2023ApJ...954..110O, Harrison2024ApJ...967...40H}. 
For AS~209, the polarization along the disk major axis has polarization angles parallel to the disk minor axis, but the polarization in the outer regions appears azimuthally oriented (at Band~7 from \citealt{Mori2019ApJ...883...16M} and at Band~6 from \citealt{Harrison2021ApJ...908..141H}). 
The polarization pattern is reminiscent of HL~Tau where the optical depth in the inner region is large while the optical depth in the outer region is low leading to a scattering signature in the inner region and a signature of toroidally aligned, prolate grains in the outer region \citep{Stephens2017ApJ...851...55S, Lin2024MNRAS.528..843L}. 

An alternative to radiative alignment is mechanical alignment where interactions with the gas can align the dust grains. One type of mechanical alignment is the Gold mechanism that allows the alignment of grains along their long axes under the presence of gas-dust relative motion \citep{Gold1952MNRAS.112..215G, Gold1952Natur.169..322G, Purcell1969Phy....41..100P}. Although the mechanism can explain the necessary prolate grains, the required relative motion should be supersonic which does not apply to protoplanetary disks \citep[e.g.][]{Nakagawa1986Icar...67..375N, Lesur2023ASPC..534..465L}. 

Another type of mechanical alignment is through the helicity of grains, where grains spin up under the presence of gas-dust relative motion \citep{Lazarian2007ApJ...669L..77L}. This mechanism also requires internal alignment, which, along with external alignment, ensures that the short axes of grains become aligned to the direction of the gas flow or the magnetic field $\vect{B}$. While the proposed mechanism no longer requires supersonic drift, it is expected to produce effectively oblate grains because of internal alignment. The observation of prolate grains suggests a lack of internal alignment which may not be too surprising since internal alignment timescales are much longer than the gas damping timescale for grains larger than $\sim 10$~$\mu$m in protoplanetary disk environments \citep{Hoang2022AJ....164..248H}. 

A potential solution is through a modification to the radiative alignment paradigm using the so-called ``wrong'' internal alignment \citep{Hoang2009ApJ...697.1316H}. For large grains without internal alignment, the short axes can become perpendicular to $\vect{J}$ (in contrast to the ``right'' alignment mentioned above where the short axes are parallel to $\vect{J}$). When $\vect{J}$ becomes externally aligned to the expected toroidal $\vect{B}$-field of a protoplanetary disk, it can match the desired toroidally aligned prolate grains if grains contain large amounts of iron clusters \citep{Thang2024ApJ...970..114T}. However, alignment with the $\vect{B}$-field appears difficult in disk environments because the Larmor precession timescale can be much longer than the gas damping timescale unless the former is greatly shortened by, e.g., superparamagnetic inclusions \citep{Tazaki2017ApJ...839...56T, Yang2021ApJ...911..125Y}.

The various examples from observations call for an answer to why grains are toroidally aligned and effectively prolate, but a robust explanation remains inconclusive. Understanding the grain alignment mechanism is not only interesting in its own right, but it will also permit the measurement of the underlying grain-aligning vector field in the disk. In this paper, we investigate a novel type of mechanical alignment, where the offset between the center of mass of a grain and its geometric center creates restoring torques when the grain feels a systematic flow of gas with respect to its center of mass, i.e., an aerodynamic flow (denoted by the vector ``$\vect{A}$'' hereafter). The mechanism is motivated by how a badminton birdie flying through the air oscillates and quickly aligns aerodynamically along the direction of flight against the gaseous headwind in terrestrial environments
\footnote{Other examples from human activities include the ji\`anzi or the hago (in the game of hanetsuki). 
}
\citep{Cohen2015NJPh...17f3001C}. 
To highlight the central role of the offset between geometric and mass centers, we will term the mechanism ``badminton birdie-like alignment" (or ``birdie-like alignment" for short) to distinguish it from other forms of mechanical alignment (the expected degree of the offset will be discussed below).
We ignore internal relaxation since the effect is slow for large grains and only consider the rotational motion of grains due to torques provided by gas drag. In disks with subsonic gas-dust bulk relative motion, we will show that the proposed mechanism can achieve fast alignment within an orbital time for most regions in disks, which opens a new window on mapping the field of aerodynamic flow acting on grains --- the $\vect{A}$-field --- through dust continuum polarization observations.

The paper is organized as follows. Section~\ref{sec:doublesphere} builds the physical intuition by using a simple grain model composed of two spheres to illustrate the key alignment behavior and alignment timescales under the presence of a gas flow. Section~\ref{sec:spheroid} derives the torque on a grain of arbitrary shape and calculates the alignment of spheroids (both prolate and oblate) whose solutions for polarization are known enabling connections to observed disk polarization from aligned grains. Section~\ref{sec:disk} implements the spheroid alignment model in a simple disk model with gas and dust velocity fields to assess the alignment timescale and alignment directions. Section~\ref{sec:discussion} provides a discussion, and we summarize our results in Section~\ref{sec:conclusion}. 

\section{Double-Sphere Illustration} \label{sec:doublesphere}

This section begins with the simplest model to illustrate the alignment under gas drag using a double-sphere model \citep{Cohen2015NJPh...17f3001C}. Given the simple spherical shape, the gas drag in the Epstein regime has a simple analytical form \citep{Epstein1924PhRv...23..710E}. We first describe the necessary formulation of the problem, and then parameterize the model to focus on the physical quantities that determine alignment. We will show that the asymmetry between the two spheres induces oscillation due to a systematic gas-dust relative motion. The spin of the grain through gas leads to damping of the oscillation and eventually to alignment. 

\subsection{Problem Setup}

\begin{figure}
    \centering
    \includegraphics[width=\columnwidth]{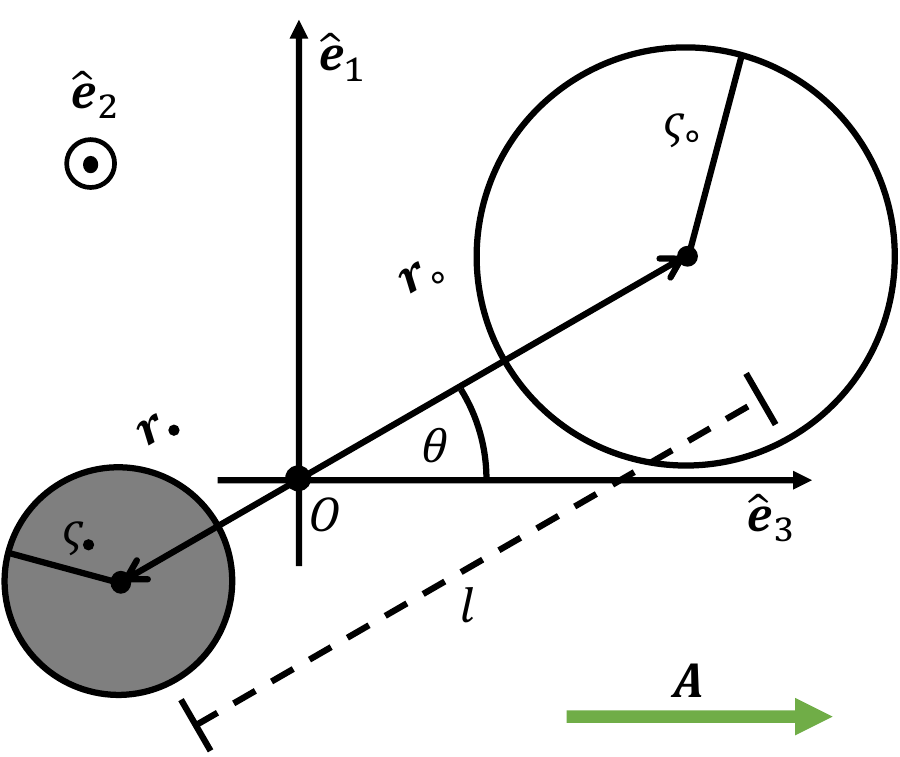}
    \caption{
        Schematic of the double sphere model. 
        The dust grain is composed of two spheres, labeled ``black'' (circle with a dark shade) and ``white'' (circle without a shade), with radii $\varsigma_{\bullet}$ and $\varsigma_{\circ}$, respectively. 
        Point $O$ is the center of mass of the grain. 
        The radius vectors from the center of mass to the center of each sphere are $\vect{r}_{\bullet}$ and $\vect{r}_{\circ}$, respectively. 
        The distance between the centers of the spheres is $l$ denoted by a dashed line.
        $\vect{A}$ is the gas velocity relative to the center of mass of the grain, i.e., the aerodynamic flow from the perspective of the grain, and is along $\vect{\hat{e}}_{3}$. 
        The spin of the grain is along $\vect{\hat{e}}_{2}$. $\theta$ is the angle of $\vect{r}_{\circ}$ from $\vect{\hat{e}}_{3}$ in the counter-clockwise direction in this figure. 
        Since alignment occurs at $\theta=0$ when $m_{\bullet}/m_{\circ} > \varsigma_{\bullet}^{2} / \varsigma_{\circ}^{2}$, the black sphere has a larger mass and is smaller than the white sphere. 
    }
    \label{fig:doublesphere_schem}
\end{figure}

Consider a grain composed of two spheres connected by a massless rigid pole. 
We use point $O$ to denote the center of mass and let $\vect{A}$ be the systematic velocity of the gas with respect to $O$. Fig.~\ref{fig:doublesphere_schem} shows a schematic of the double-sphere.
To distinguish each sphere, we use subscripts ``$\bullet$'' and ``$\circ$'' to represent the black and white spheres, respectively. 
The radius vectors from $O$ to the center of each sphere are $\vect{r}_{\bullet}$ and $\vect{r}_{\circ}$, respectively, with corresponding lengths denoted as $r_{\bullet}$ and $r_{\circ}$. 
By definition of the center of mass, the lengths are related to the mass through
\begin{align} \label{eq:center_of_mass}
    m_{\bullet} r_{\bullet} = m_{\circ} r_{\circ} \text{ ,}
\end{align}
where $m_{\bullet}$ and $m_{\circ}$ are the masses of the spheres. The $i$th sphere ($i=\bullet$ or $\circ$) is characterized by radius $\varsigma_{i}$ and its radius and mass are related through 
\begin{align}
    m_{i} = \frac{4\pi}{3} \varsigma_{i}^{3} \rho_{i} \text{ ,}
\end{align}
where $\rho_{i}$ is the material density for the $i$th sphere, which can be different for the two spheres.

We use the Cartesian coordinate system centered on the center of mass $O$ with unit directions $\vect{\hat{e}}_{1}$, $\vect{\hat{e}}_{2}$, and $\vect{\hat{e}}_{3}$. Let $\vect{\hat{e}}_{3}$ be along the direction of the gas flow $\vect{A}$. 
We define the angle between $\vect{r}_{\circ}$ and $\vect{\hat{e}}_{3}$ as $\theta$. 
Without loss of generality, 
let $\vect{\hat{e}}_{1}$ be in the plane formed by $\vect{r}_{\circ}$ and $\vect{\hat{e}}_{3}$. The aerodynamic drag force will be in this plane, producing a torque around the $\vect{\hat{e}}_{2}$ axis, which will be the axis of the flow-induced grain rotating motion, denoted by $\vect{\omega} = \omega \vect{\hat{e}}_{2}$.

The drag force in the Epstein regime is \citep{Epstein1924PhRv...23..710E}: 
\begin{align} \label{eq:epstein_drag}
    \vect{F} = - \frac{4 \pi }{3} \rho_{g} \varsigma^{2} \vth \vect{u}
\end{align}
for a spherical grain of radius $\varsigma$ embedded in a gas with mass density $\rho_{g}$. For convenience, we will also use the gas number density $n_{g} \equiv \rho_{g} / (\mu m_{p})$ where $m_{p}$ is the mass of a proton and $\mu$ is the mean molecule weight which we assume as 2.3 for molecular hydrogen-dominated gas. 
$\vth$ is the average speed of the gas molecules with a Maxwell-Boltzmann distribution
\begin{align}
    \vth \equiv \sqrt{\frac{8 k T}{\pi \mu m_{p}}}
\end{align}
where $k$ is the Boltzmann constant and $T$ is the gas temperature. 
$\vect{u}$ is the velocity of the sphere relative to the gas
\footnote{There are several quantities related to velocity used throughout the paper that deserve clarification. We use ``$A$'' to denote the bulk motion of gas relative to the center of mass of a grain. We use ``$u$'' to denote the motion of a \textit{piece} of a grain relative to the gas. For Sec.~\ref{sec:doublesphere}, $u$ is the speed for one of the spheres, while for Sec.~\ref{sec:spheroid}, $u$ corresponds to the speed of an infinitesimal surface. In Sec.~\ref{sec:disk}, we will use ``$v$'' to describe the velocity field of the disk.
}. 
The Epstein regime applies if the mean free path of the gas molecules is much larger than the size of the particle. For disks, the mean free path is generally of order $\sim$cm near $\sim 1$~au and larger at larger radii where the gas density is lower \citep{Armitage2015arXiv150906382A}. Thus, the Epstein regime is appropriate in the outer disk where resolved dust polarization is detected (see Sec.~\ref{sec:disk_alignment_timescale} for an estimate of the mean free path).


Each sphere provides a cross-section for gas drag to induce a torque $\vect{\Gamma}_{i} \equiv \vect{r}_{i} \times \vect{F}_{i}$ on the grain. 
The velocity $\vect{u}_{i}$ for the $i$th sphere traveling relative to the gas is
\begin{align} \label{eq:doublesphere_velocity_difference}
    \vect{u}_{i} \equiv \vect{\omega} \times \vect{r}_{i} - \vect{A}
\end{align}
when incorporating the rotation of the grain (with angular velocity $\vect{\omega}$). 
Applying the relative velocity from Eq.~\ref{eq:doublesphere_velocity_difference} to the drag force from Eq.~\ref{eq:epstein_drag} gives 
\begin{align} \label{eq:doublesphere_torque_vector}
    \vect{\Gamma}_{i} = - \frac{4\pi}{3} \rho_{g} \vth \varsigma_{i}^{2} [ \vect{r}_{i} \times (\vect{\omega} \times \vect{r}_{i}) - \vect{r}_{i} \times \vect{A} ]. 
\end{align}
The total torque on the grain is $\vect{\Gamma} = \vect{\Gamma}_{\bullet} + \vect{\Gamma}_{\circ}$ from both spheres. 

We can easily identify that the torque $\vect{\Gamma}$ is only non-zero along $\vect{\hat{e}}_{2}$ meaning we only have to consider the component $\Gamma_{2}$ ($\Gamma_{1}=\Gamma_{3}=0$). 
There is only one equation of motion: 
\begin{align} \label{eq:torque_equals_I_omega_double_sphere}
    \Gamma_{2} = I \dot{\omega}
\end{align}
where $I$ is the moment of inertia and $\dot{\omega}$ is the angular acceleration. For the double-sphere model, we can easily derive the moment of inertia as 
\begin{align} \label{eq:doublesphere_moment_of_inertia}
    I = \sum_{i} m_{i} \bigg(\frac{2}{5} \varsigma_{i}^{2} + r_{i}^{2}\bigg). 
\end{align}
where the second term comes from the parallel axis theorem. 

Using Eq.~\ref{eq:doublesphere_torque_vector} and \ref{eq:torque_equals_I_omega_double_sphere}, we get a second-order differential equation: 
\begin{align} \label{eq:equation_of_motion}
    I \dot{\omega} + D \omega + P \sin \theta &= 0
\end{align}
where 
\begin{align}
    D &\equiv \frac{4\pi}{3} \rho_{g} \vth (\varsigma_{\bullet}^{2} r_{\bullet}^{2} + \varsigma_{\circ}^{2} r_{\circ}^{2}) \nonumber \\
    P &\equiv \frac{4\pi}{3} \rho_{g} \vth A (- \varsigma_{\bullet}^{2} r_{\bullet} + \varsigma_{\circ}^{2} r_{\circ}). \nonumber
\end{align} 
The second term is a torque that is proportional to $\omega$ and acts to oppose it, serving as the damping torque. The third term is a $\theta$-dependent torque induced by the gas flow through relative dust-gas drift.

In the limit of small $\theta$, Eq.~\ref{eq:equation_of_motion} simply describes a damped harmonic oscillator. 
The motion of a damped harmonic oscillator is characterized by two timescales: the undamped period of oscillation $t_{o}$ and the damping time $t_{d}$. The undamped period of oscillation is simply 
\begin{align}
    t_{o} \equiv \frac{2 \pi }{\omega_{o}}
\end{align}
where $\omega_{o}$ is the undamped angular frequency
\begin{align} \label{eq:undamped_angular_frequency}
    \omega_{o} \equiv \sqrt{ \frac{P}{I} }. 
\end{align}
The damping time describes the timescale for the oscillation amplitude to decrease:
\begin{align} \label{eq:damping_time}
    t_{d} \equiv \frac{2 I}{D}. 
\end{align}
Although these expressions are derived from analytical solutions to the damped harmonic oscillator, it remains beneficial to use them as characteristic timescales for Eq.~\ref{eq:equation_of_motion} beyond the limit of small $\theta$. Evidently, with the birdie-like alignment mechanism, the timescale to reach alignment is in fact the damping time, which we discuss in more detail below. 

\subsection{Alignment Direction}

\begin{figure*}
    \centering
    \includegraphics[width=0.9\textwidth]{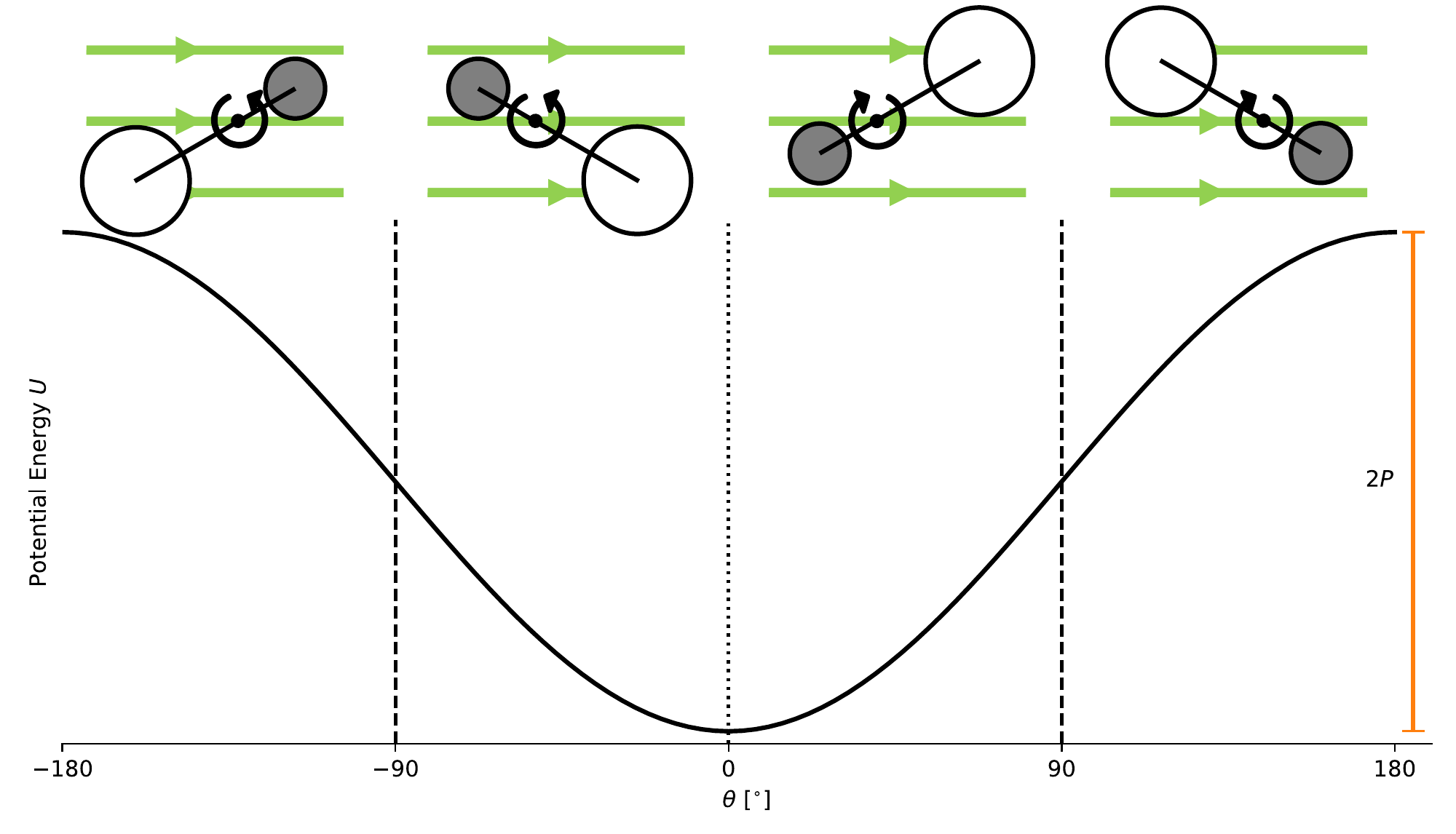}
    \caption{
        A schematic of how the orientation of the grain relates to the potential well created by the existence of aerodynamic flow. Top panel: the orientation of an example double-sphere grain model that satisfies $m_{\bullet}/m_{\circ} > \varsigma_{\bullet}^{2} / \varsigma_{\circ}^{2}$ like in Fig.~\ref{fig:doublesphere_schem} The heavier sphere is colored gray, while the less massive sphere is white. The green lines with arrows denote the direction of the gas flow. The circular arrow depicts the direction of the flow-induced torque, which always tries to align the grain and resists any angular displacement. As such, the flow-induced torque is a ``restoring" torque. 
        When aligned, the less massive sphere follows the direction of the flow, analogous to how the heavier head of the badminton birdie leads the less massive tail against a headwind. 
        Bottom panel: the energy potential $U$ as a function of $\theta$. The depth of the potential well is $2P$. Alignment occurs at $\theta=0$ where $U$ is minimal. 
    }
    \label{fig:potential_schematic}
\end{figure*}

When $P\neq 0$ (i.e., when the two grains are not identical and $A \neq 0$), we can easily understand why this grain has to be aligned by defining the potential energy from the $\theta$-dependent torque:
\begin{align} \label{eq:potential_energy}
    U \equiv - \int_{0}^{\theta} (-P \sin \theta') d \theta' = - P \cos \theta
\end{align}
where $\theta'$ is a dummy variable and the reference potential energy at $\theta=0$ is $-P$. 
We first consider $P > 0$, in which case $U$ is a minimum at $\theta=0$. 
The $- \cos \theta$ dependence of $U$ is easy to understand through Fig.~\ref{fig:potential_schematic} which shows a schematic of how the orientation of the grain relates to $U$ described by Eq.~\ref{eq:potential_energy}. 
The direction of the $\theta$-dependent torque depends on the orientation of the grain and acts against the displacement of $\theta$ from $0$ attempting to trap the grain in alignment. At $\theta=0$, the torque vanishes and the grain is kept stable with the white sphere following the direction of the flow (the less massive and/or larger sphere follows $\vect{A}$). 
The anti-alignment point, $\theta=\pi$, also does not produce a torque, but is unstable. The $\theta$-dependent torque is thus a restoring torque within the potential well, which we call the ``flow-induced restoring torque.''

The existence of the damping torque ($\omega$-dependent torque) diminishes the spin, which eventually decreases the rotational energy until $\theta=0$, i.e., the grain becomes aligned, which corresponds to the minimum energy state. 
Alternatively, if $P<0$ (for example, the white sphere becomes the heavier sphere), then $U$ reaches a minimum at $\theta=\pi$ though the less massive and/or larger sphere still follows the direction of the flow (recall that $\theta$ is specifically defined using $\vect{r}_{\circ}$; see Fig.~\ref{fig:doublesphere_schem}). 

From Eq.~\ref{eq:potential_energy}, the peak of the potential well is simply $2P$. If the total energy (rotational plus potential energy) of a grain is greater than $2P$, the grain will not be trapped in a potential well, but continue to spin (and not oscillate). For an initially aligned grain, one can easily derive that the escape angular speed is $\omega_{\text{esc}} = 2 \omega_{o}$, i.e., a grain spinning faster than $\omega_{\text{esc}}$ cannot be trapped in oscillation. As gas damping dissipates the rotational energy of the grain, the grain will inevitably land in a potential well and begin oscillating until it reaches alignment at $\theta=0$.

Observing the expression of $P$ from Eq.~\ref{eq:equation_of_motion}, we can realize that the quantity $- \varsigma_{\bullet}^{2} r_{\bullet} + \varsigma_{\circ}^{2} r_{\circ}$ is the first moment of the cross-sections of the spheres. We can define the ``geometrical center'' of the double-sphere grain by
\begin{align} \label{eq:double_sphere_geometrical_center}
    r_{g} \equiv \frac{- \varsigma_{\bullet}^{2} r_{\bullet} + \varsigma_{\circ}^{2} r_{\circ}}{\varsigma_{\bullet}^{2} + \varsigma_{\circ}^{2}}
\end{align}
and use $\vect{r}_{g}$ to denote the vector from the center of mass to the geometrical center.
When $r_{g}>0$, the geometrical center shifts towards the white sphere and the final alignment angle at $\theta=0$ is such that $\vect{r}_{g} \parallel \vect{A}$. 
Conversely, when $r_{g}<0$, the geometrical center moves closer to the black sphere, and the final alignment angle at $\theta=\pi$ also means $\vect{r}_{g} \parallel \vect{A}$. 
In other words, the asymmetry between the two spheres causes an offset between the center of mass and the geometrical center and the grain becomes susceptible to alignment with $\vect{A}$. 
The final alignment orientation is such that the geometrical center points along the direction of the flow relative to the center of mass. 
When $r_{g}=0$, the geometrical center corresponds to the center of mass leading to $P=0$. 

Since the center of mass is related to the mass of the spheres, we can express how alignment depends on the mass or density.
To fulfill $P>0$, it requires 
\begin{align}
    \frac{m_{\bullet}}{ m_{\circ} } > \frac{ \varsigma_{\bullet}^{2} }{ \varsigma_{\circ}^{2} }
\end{align}
where we used the definition of $P$ and Eq.~\ref{eq:center_of_mass}. If the spheres have equal mass, $m_{\bullet} = m_{\circ}$, then the black sphere should be smaller than the white sphere ($\varsigma_{\circ} > \varsigma_{\bullet}$) to reach alignment at $\theta=0$. 

Equivalently, we can re-express $P>0$ using the material density and obtain 
\begin{align} \label{eq:rho_ratio_a_ratio_for_P}
    \frac{\rho_{\bullet}}{\rho_{\circ}} \frac{\varsigma_{\bullet}}{\varsigma_{\circ}} > 1. 
\end{align}
This expression explains that both an asymmetry in the material density and/or in the size can create alignment. 
If $\rho_{\bullet} = \rho_{\circ}$, the black sphere needs to be larger than the white sphere $\varsigma_{\bullet} > \varsigma_{\circ}$ to reach alignment at $\theta=0$ (note the opposite relation because the material density is fixed as opposed to keeping the total mass fixed). 

Lastly, we consider the behavior when the flow-induced torque is non-existent.
From Eq.~\ref{eq:equation_of_motion}, we can see that $P$ can equal 0 when the two spheres are identical or when $A=0$, while $D$ is always greater than 0. When $P=0$, the equation reduces to $I \dot{\omega} + D \omega = 0$, which can be integrated once to yield the time evolution of the grain misalignment angle relative to the flow:
\begin{align} \label{eq:solution_P_eq_0}
    \theta(t) = \theta(0) + \frac{1}{2} \omega(0) t_{d} ( 1 - e^{- 2 t / t_{d}} )
\end{align}
where $\theta(0)$ and $\omega(0)$ are the initial conditions of $\theta$ and $\omega$. 
Eq.~\ref{eq:solution_P_eq_0} means that the grain does not oscillate. 
The final orientation of $\theta$($t \rightarrow \infty$) is $\theta(0) + \omega(0) t_{d} / 2$. 
In other words, without the restoring torque ($P=0$), the final alignment angle depends on the initial conditions. 
If the initial conditions for the grains are random \textit{and} the direction of rotation is also random, then the ensemble should not have any preferred alignment direction (see Section~\ref{sec:discussion} for a discussion on the resulting polarization).

\subsection{Parameterization}

While the above discussion allows a qualitative description of alignment, we utilize a parameterization of the double-sphere model to facilitate the quantitative description. In particular, we aim to quantify how $t_{o}$ and $t_{d}$ depend on the characteristic properties of the grain, in addition to the level of asymmetry between the spheres. We define the characteristic length $l$ of the entire grain through
\begin{align}
    l \equiv r_{\bullet} + r_{\circ}
\end{align}
which is simply the length between the center of the two spheres. We define the ratio of the sizes through
\begin{align} \label{eq:radius_ratio}
    \epsilon \equiv \frac{\varsigma_{\bullet}}{\varsigma_{\circ}}
\end{align}
and the sum of the sizes is constrained by 
\begin{align}
    \varsigma_{\bullet} + \varsigma_{\circ} \equiv \lambda l
\end{align}
where $\lambda$ is the fraction of $l$ that the radius of each sphere occupies with $\lambda \leq 1$. $\lambda=1$ means the two spheres are in contact. 
The sizes are thus
\begin{align}
    \varsigma_{\bullet} &= \frac{ \epsilon \lambda }{ 1 + \epsilon } l, \nonumber \\
    \varsigma_{\circ} &= \frac{ \lambda }{ 1 + \epsilon } l.
\end{align}

In addition to the size, we define the ratio of the densities through
\begin{align} \label{eq:density_ratio}
    \kappa \equiv \frac{ \rho_{\bullet} }{ \rho_{\circ} }. 
\end{align}
We denote the total mass of the grain through $m \equiv m_{\bullet} + m_{\circ}$ and along with definitions of $\epsilon$ and $\kappa$, one can find that
\begin{align}
    m_{\bullet} &= \frac{ \epsilon^{3} \kappa }{ 1 + \epsilon^{3} \kappa } m ,  \nonumber \\
    m_{\circ} &= \frac{ 1 }{ 1 + \epsilon^{3} \kappa } m. 
\end{align}
Since $r_{i}$ for each sphere is related to the center of mass through Eq.~\ref{eq:center_of_mass}, we can re-express $r_{i}$ by 
\begin{align}
    r_{\bullet} &= \frac{1}{1 + \epsilon^{3} \kappa } l , \nonumber \\
    r_{\circ} &= \frac{\epsilon^{3} \kappa}{1 + \epsilon^{3} \kappa } l.
\end{align}

It is also convenient to consider the overall material density $\rho_{s}$ which is the total mass encompassed by the total volume of the two spheres: 
\begin{align}
    \rho_{s} = \frac{m}{l^{3}} \frac{3}{4\pi \lambda^{3}} \frac{(1 + \epsilon)^{3}}{1 + \epsilon^{3}}
\end{align}
One can find that 
\begin{align}
    \rho_{\bullet} &= \frac{ 1 + \epsilon^{3} }{ 1 + \epsilon^{3} \kappa } \kappa \rho_{s} \nonumber \\
    \rho_{\circ} &= \frac{ 1 + \epsilon^{3} }{ 1 + \epsilon^{3} \kappa } \rho_{s}
\end{align}

We can now determine the coefficients to Eq.~\ref{eq:equation_of_motion}. The moment of inertia is 
\begin{align}
    I = \frac{4\pi}{3} \rho_{s} l^{5} \frac{ \lambda^{3} ( 1 + \epsilon^{3} ) }{(1 + \epsilon)^{3} (1 + \epsilon^{3} \kappa)}
        \bigg[ 
            \frac{\epsilon^{3} \kappa}{1 + \epsilon^{3} \kappa} 
            + \frac{2}{5} \frac{ \lambda^{2} (1 + \epsilon^{5} \kappa ) }{ (1 + \epsilon)^{2} }
        \bigg]
\end{align}
The coefficient to the drag term is 
\begin{align}
    D = \frac{4\pi}{3} \rho_{g} \vth l^{4} \frac{\lambda^{2} \epsilon^{2} (1 + \epsilon^{4} \kappa^{2})}{(1 + \epsilon)^2 (1 + \epsilon^{3} \kappa)^2}
\end{align}
Lastly, we find 
\begin{align}
    P = \frac{4\pi}{3} \rho_{g} \vth A l^{3} \frac{ \lambda^{2} \epsilon^{2} ( \epsilon \kappa - 1) }{ (1 + \epsilon)^2 (1 + \epsilon^{3} \kappa)}. 
\end{align}
Note that the factor $\epsilon \kappa - 1$ is equivalent to Eq.~\ref{eq:rho_ratio_a_ratio_for_P} and determines the sign of $P$, which, in turn, determines the alignment direction.

\begin{figure}
    \centering
    \includegraphics[width=\columnwidth]{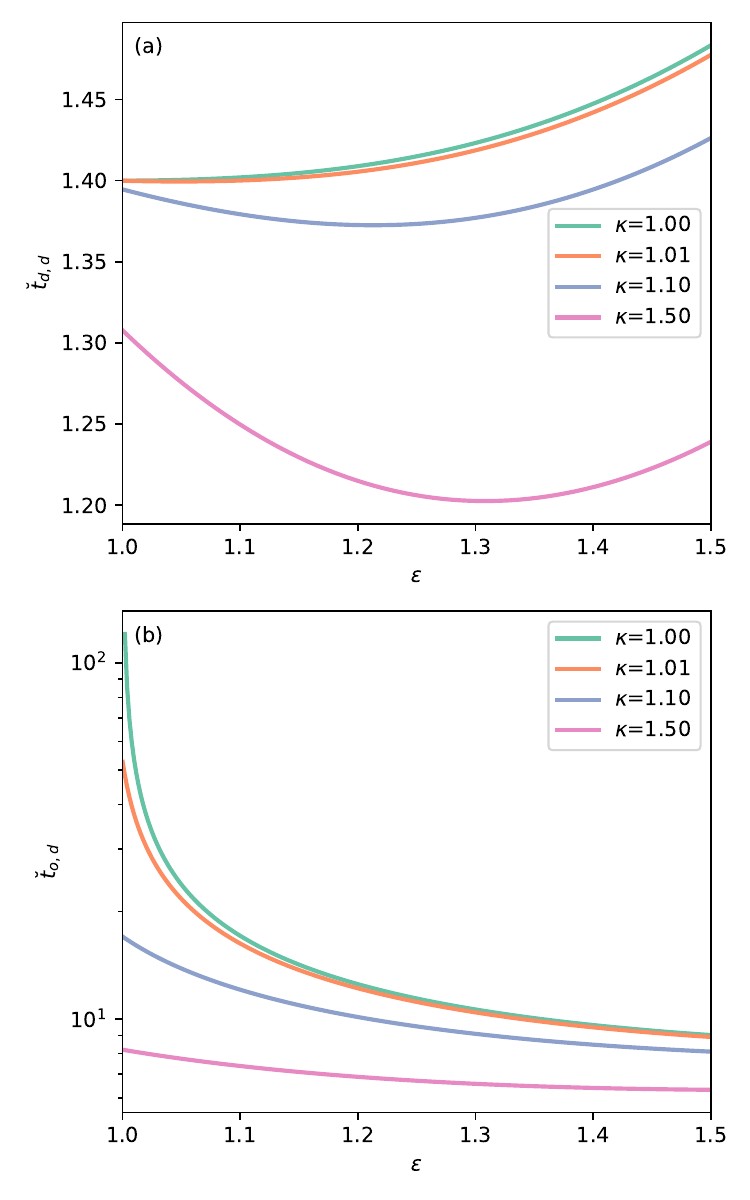}
    \caption{
        Dimensionless factors for the double-sphere model as a function of $\epsilon$ (ratio of radii between the two spheres) and $\kappa$ (ratio of material density). Panel a: the damping time dimensionless factor, $\Breve{t}_{d,d}$. Panel b: the oscillation time dimensionless factor, $\Breve{t}_{o,d}$. As $\epsilon\rightarrow1$ and $\kappa\rightarrow1$, $\Breve{t}_{o,d} \rightarrow \infty$ meaning there is no oscillation when the two spheres are equivalent. 
    }
    \label{fig:doublesphere_time_factor}
\end{figure}

With the coefficient determined, we can assess the damping time $t_{d}$ and oscillation time $t_{o}$. We will use $t_{d,d}$ and $t_{o,d}$ to denote $t_{d}$ and $t_{o}$ of the double-sphere model, respectively. 
The damping time for the double-sphere is 
\begin{align}
    t_{d, d} &= \frac{\rho_{s} l}{\rho_{g} \vth} \Breve{t}_{d,d} \text{ ,} \nonumber \\
    \Breve{t}_{d,d} &\equiv 
    \frac{
        2 \lambda (1+\epsilon^{3}) (1 + \epsilon^{3} \kappa)
    }{
        \epsilon^{2} (1 + \epsilon) (1 + \epsilon^{4} \kappa^{2})
    } 
    \bigg[ 
            \frac{\epsilon^{3} \kappa}{1 + \epsilon^{3} \kappa} 
            + \frac{2}{5} \frac{ \lambda^{2} (1 + \epsilon^{5} \kappa ) }{ (1 + \epsilon)^{2} }
        \bigg]
\end{align}
where $\Breve{t}_{d,d}$ is a dimensionless factor that encapsulates the asymmetry. 
We will use a breve ( $\Breve{ }$ ) to denote the dimensionless factors that only depend on the geometry of the grain. 
The physical quantity $\rho_{s} l / (\rho_{g} \vth)$ is equivalent to the stopping time of a spherical grain with radius $l$ and material density $\rho_{s}$. 
The stopping time is the characteristic timescale for the aerodynamic drag force to stop the motion of a moving grain \citep{Armitage2015arXiv150906382A}. 
The relation between the damping time and the stopping time of a single sphere is not too surprising. As the double-sphere grain spins, each sphere moves relative to the gas. 
Requiring each sphere to stop its relative motion with respect to the gas is equivalent to stopping the spin of the system, i.e., rotational damping. 
Thus, the damping time should be on the order of the stopping time. However, this effect cannot be captured for a single spherical grain as we will see in Section~\ref{sec:spheroid}. 
Plugging in some typical numbers for a protoplanetary disk for the stopping time, we get 
\begin{align} \label{eq:double_sphere_stopping_time}
    t_{s} 
    &\equiv \frac{ \rho_{s} l }{ \rho_{g} \vth } \nonumber \\ 
    &\sim 25 \text{ years } \bigg( \frac{ \rho_{s} }{ 3 \text{ g cm$^{-3}$ } } \bigg) 
        \bigg( \frac{ l }{ 1 \text{ mm } }\bigg) 
        \bigg( \frac{ 10^{9} \text{ cm$^{-3}$ } }{ n_{g} } \bigg) 
        \bigg( \frac{ 1 \text{ km/s } }{ \vth } \bigg) .
\end{align}
Since increasing gas density dampens the grain spin more quickly ($t_{d} \propto 1 / \rho_{g}$), we expect that grains are better aligned in higher gas densities. 
Also, since $t_{d} \propto l$, larger grains are more difficult to be damped and should be less aligned.

Fig.~\ref{fig:doublesphere_time_factor}a shows the asymmetry factor $\Breve{t}_{d,d}$ for different values of size ratio $\epsilon$ and density ratio $\kappa$ while adopting $\lambda=1$ (i.e., contacting spheres). 
We find that the value is of order unity across the parameter space. When $\epsilon=1$ and $\kappa=1$ (the symmetric case), then $\Breve{t}_{d,d}=1.4$. Note that the flow speed $A$ does not contribute to the damping time, but only determines the oscillation potential (and hence frequency) around the dust-gas drift direction.

The oscillation time for the double-sphere is 
\begin{align}
    t_{o, d} &= \sqrt{ \frac{\rho_{s} l^{2}}{\rho_{g} \vth A} } \Breve{t}_{o,d} \text{ ,}  \nonumber \\
    \Breve{t}_{o,d} &\equiv 2 \pi 
    \sqrt{ \frac{ \lambda ( 1 + \epsilon^{3} ) }{ \epsilon^{2} (1 + \epsilon)  ( \epsilon \kappa - 1) }
        \bigg[ 
            \frac{\epsilon^{3} \kappa}{1 + \epsilon^{3} \kappa} 
            + \frac{2}{5} \frac{ \lambda^{2} (1 + \epsilon^{5} \kappa ) }{ (1 + \epsilon)^{2} }
        \bigg] }
\end{align}
where $\Breve{t}_{o,d}$ is a dimensionless factor for the oscillation time and the physical quantity in front of it is the characteristic oscillation time. Plugging in the same numbers for a protoplanetary disk as Eq.~\ref{eq:double_sphere_stopping_time} and adopting $A=10$~m/s (see Sec.~\ref{sec:disk} below), the characteristic oscillation time is $\sim 5$~minutes. 

Fig.~\ref{fig:doublesphere_time_factor}b shows $\Breve{t}_{o,d}$ as a function of size ratio $\epsilon$ for different values of density ratio $\kappa$. While most of the parameter space shows $\Breve{t}_{o,d}\sim 10$, its value increases rapidly as $\epsilon$ and $\kappa$ decreases to 1. Nevertheless, even if $\kappa=1.01$ and $\epsilon=1$, or $\kappa=1$ and $\epsilon=1.01$, $\Breve{t}_{o,d} \sim 50$. Thus, $t_{o,d}$ is much shorter than the damping time and the Keplerian orbital periods of the outer disk even for $1\%$ asymmetry. 

Since $t_{o,d} \ll t_{d,d}$, the grain undergoes many oscillations before the amplitude of oscillation diminishes. 
The oscillation behavior is more akin to an underdamped harmonic oscillator which enters its equilibrium state (i.e., becomes aligned) gradually. 
We can understand why through the damping ratio defined by $\zeta \equiv D / (2 \sqrt{IP})$ where $\zeta < 1$, $\zeta=1$, and $\zeta>1$ correspond to underdamped, critically damped, and overdamped oscillators, respectively. 
Applying the constants from Eq.~\ref{eq:equation_of_motion}, we have
\begin{align} \label{eq:damping_ratio}
    \zeta_{d} &= \sqrt{ \frac{\rho_{g} \vth}{\rho_{s} A} } \Breve{\zeta}_{d} 
        \text{ , } \nonumber \\
    \Breve{\zeta}_{d} &\equiv 
        \frac{ \epsilon ( 1 + \epsilon^{4} \kappa^{2}) }{2 (1 + \epsilon^{3} \kappa) }
        \sqrt{ 
            \frac{ (1 + \epsilon)^{3} }{ \lambda (1 + \epsilon^{3}) (\epsilon \kappa - 1) }  
        }
        \bigg[ 
                \frac{\epsilon^{3} \kappa}{1 + \epsilon^{3} \kappa} 
                + \frac{2}{5} \frac{ \lambda^{2} (1 + \epsilon^{5} \kappa ) }{ (1 + \epsilon)^{2} }
        \bigg]^{-\frac{1}{2}}
\end{align}
where $\Breve{\zeta}_{d}$ encapsulates the dimensionless, asymmetry factors. 
Using typical conditions of disks used in Eq.~\ref{eq:double_sphere_stopping_time}, we have $\zeta_{d} \sim 4 \times 10^{-7} \Breve{\zeta}_{d}$ and $\Breve{\zeta}_{d}\sim 6$ for $\epsilon=1.01$ and $\kappa=1$. 
One can see that $\zeta_{d} \ll 1$ because the gas mass density $\rho_{g}$ is drastically smaller than grain material density $\rho_{s}$ (by 15 orders of magnitude with $n_{g}=10^{9}$~cm$^{-3}$), which is more than enough to compensate for the difference between the dust-gas relative speed $A$ and the thermal speed $\vth$.

\begin{figure*}
    \centering
    \includegraphics[width=0.9\textwidth]{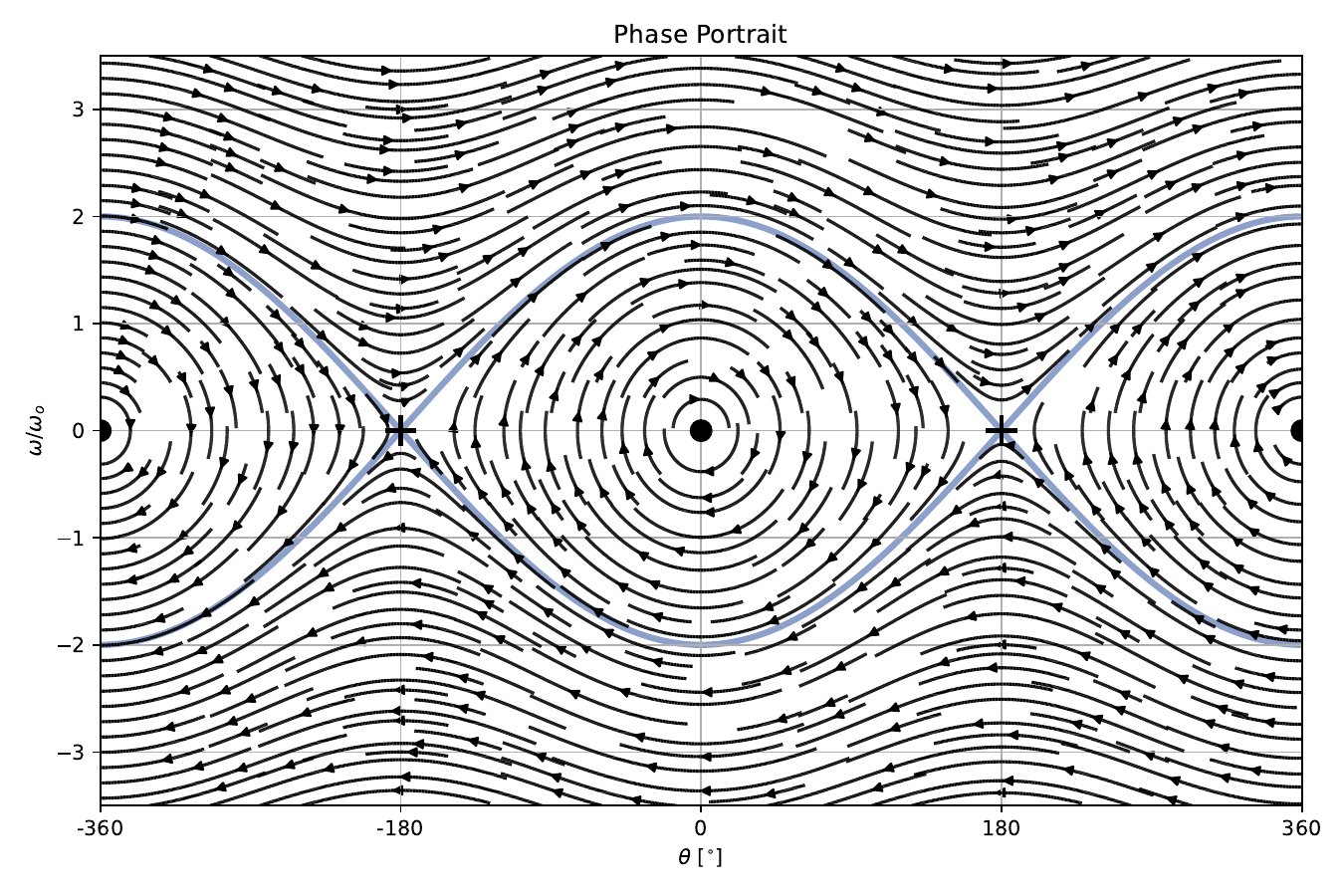}
    \caption{
        The phase portrait (trajectory map) of the double-sphere grain under the presence of gas flow $\vect{A}$. The horizontal axis is $\theta$, while the vertical axis is the angular velocity $\omega$ normalized by the oscillation frequency $\omega_{o}$. The black circles correspond to stable equilibrium points that serve as the attractors for alignment. The plus signs are the unstable equilibrium points. The blue curve is the separatrix that separates the bounded and unbounded regions. Regions inside the separatrix are bounded and oscillate around the attractors. 
    }
    \label{fig:doublesphere_phase_portrait}
\end{figure*}

Fig.~\ref{fig:doublesphere_phase_portrait} shows the phase portrait (or trajectory maps) of the damped oscillations given by Eq.~\ref{eq:equation_of_motion}. 
The phase portrait shows the evolution of $\theta$ and $\omega$ for the dynamical system\footnote{We use the Python package \textsc{phaseportrait} available at \url{https://phaseportrait.github.io}}. 
We used the same values from Eq.~\ref{eq:double_sphere_stopping_time} along with $\epsilon=1.01$, $\kappa=1$, and $A=10$~m/s.
Fig.~\ref{fig:doublesphere_phase_portrait} also shows the separatrix described by $(\omega/\omega_{0})^{2} = 2 ( 1 + \cos \theta)$ which separates the bounded and unbounded region. 
One can derive this relation by equating the total energy (kinetic and potential energy) of the system to the energy at the unstable point. Grains in the unbounded regions continually spin without regard to the alignment direction. 
However, as the grain loses energy through damping, it will cross the separatrix at some point and become bound within a potential well (see Appendix~\ref{sec:better_damped_phase_portrait} for an example phase portrait when the grain is more damped). 
We can see that $\theta$ of a grain oscillates around the alignment point at $\theta=0$ and is repelled at $\theta=-\pi$ and $\pi$.

To conclude the results from the double-sphere model, the asymmetry in the structure (a combination of the cross-section and mass) and the existence of gas flow lead to restoring torques that make the grain oscillate instead of a spin-up.
As the grain oscillates, the torque from gas damping opposes any grain rotating motion which dampens the magnitude of oscillation. 
Even a $1\%$ level of asymmetry allows the grain to undergo severely underdamped oscillations. 
The damped oscillations eventually reach an orientation where the grain geometric center aligns along the gas flow with respect to its center of mass.
The timescale to align the grain is related to the damping timescale which is comparable to the stopping time of the entire grain. 
In other words, the grain asymmetry and gas flow determine the direction of alignment and the damping time determines how quickly the grain reaches alignment.

%
%
\section{Spheroid Model} \label{sec:spheroid}

The double-sphere model of a grain allowed estimates of the oscillation behavior depending on the level of asymmetry and the timescales for alignment. 
In addition, the model will always be aligned along the long axis connecting the two spheres, which hints at a potential solution that can match the requirement of effectively aligned prolate grains constrained from observations. 
However, the shape of the model is too idealized to assess its polarization from thermal emission. In this section, we utilize an axisymmetric ellipsoid, i.e., spheroid (prolate and oblate), and assess the alignment behavior. 
The polarization from spheroids is well established, both in the dipole limit \citep{vandeHulst1957lssp.book.....V} and for arbitrary sizes using the T-matrix technique \citep{Waterman1971PhRvD...3..825W, Mishchenko1994OptCo.109...16M}. 
The spheroidal alignment model can be directly compared to observed polarization.

\subsection{Problem Setup}

Similar to the previous section, we aim to solve the total torque $\vect{\Gamma}$ of a spinning grain embedded in a flowing background of gas. Obtaining $\vect{\Gamma}$ will allow us to derive the equation of motion. While the double-sphere model only adds torque from two discrete points (the centers of the two spheres), a smooth object, like a spheroid, requires us to integrate the torque over its surface $S$:
\begin{align} \label{eq:torque_over_surface}
    \vect{\Gamma} = \oint_{S} \vect{r} \times d \vect{F} 
\end{align}
where $\vect{r}$ is the radius vector of a surface element $d\sigma$ from the center of mass $O$ and $d \vect{F}$ is the infinitesimal force from the gas acting on the surface element. 

Assuming specular reflection, gas drag in the Epstein regime, and subsonic relative motion, 
Appendix~\ref{sec:force_per_unit_surface_derivation} shows that $d \vect{F}$ at each point of the surface follows
\begin{align} \label{eq:force_per_unit_area}
    d \vect{F} = - \rho_{g} \bigg[ c_{s}^{2} \vect{n} + \vth \matr{N}(\vect{u}) \bigg] d \sigma
\end{align}
where $\vect{n}$ is the unit normal direction of $d \sigma$ and $\vect{u}$ is the velocity of the surface relative to the gas. $\matr{N}\equiv \vect{n} \vect{n}$ is a projection tensor that simply returns a vector from the part of $\vect{u}$ that is normal to the surface $d \sigma$. The negative sign means the force is opposite to the direction of $\vect{n}$. $c_{s}$ is the isothermal sound speed $c_{s}\equiv \sqrt{k T / (\mu m_{p})}$.

Assume the grain spins as a solid body with angular velocity $\vect{\omega}$ around the center of mass. Each point of the surface will travel relative to the gas as 
\begin{align} \label{eq:solid_body_velocity_difference}
    \vect{u} = \vect{\omega} \times \vect{r} - \vect{A}.
\end{align}
The expression is directly related to Eq.~\ref{eq:doublesphere_velocity_difference}, but in this case, the $\vect{u}$ is the relative velocity of a surface element and not a complete sphere that was assumed for the double-sphere model. 
Substituting Eq.~\ref{eq:solid_body_velocity_difference} and Eq.~\ref{eq:force_per_unit_area} into Eq.~\ref{eq:torque_over_surface} yields three terms to the total torque on the spheroid: 
\begin{align} \label{eq:torque_three_terms_with_tensor}
    \vect{\Gamma} = 
        - \rho_{g} c_{s}^{2} \vect{K} 
        + \rho_{g} \vth \matr{L} \vect{\omega} 
        + \rho_{g} \vth \matr{M} \vect{A}
\end{align}
with
\begin{align}
    \vect{K} &\equiv \oint_{S} \vect{r} \times \vect{n} d \sigma, \nonumber \\
    \matr{L} &\equiv \oint_{S} \vect{r} \times \matr{N} (\vect{r} \times) d \sigma, \nonumber \\
    \matr{M} &\equiv \oint_{S} \vect{r} \times \matr{N} d\sigma \nonumber
\end{align}
where $\vect{K}$ is a vector and $\matr{L}$ and $\matr{M}$ are tensors. We have rearranged the expression such that only quantities that depend on the location of the surface are kept inside each integral. Note that the cross product, $\vect{r} \times$, is a 2-rank tensor that can be implemented as a matrix calculation. 
The first term is a driving torque that exists regardless of any $\vect{A}$ or $\vect{\omega}$. 
The second term is the damping torque since it is related to $\vect{\omega}$. 
The third term depends on $\vect{A}$ and, as we will see later, produces the restoring torque.

Eq.~\ref{eq:torque_three_terms_with_tensor} is particularly useful since each integral only depends on the structure of the grain and does not depend on the environmental properties, like $\vect{A}$ or $\rho_{g}$, or its dynamic state, like $\theta$ or $\vect{\omega}$. This opens the possibility of precalculating these integrals for any grain as its unique physical property before considering its dynamic behavior.

\subsection{Strictly Axisymmteric Spheroid} \label{sec:strictly_axisymmetric_spheroid}
\begin{figure*}
    \centering
    \includegraphics[width=0.8\textwidth]{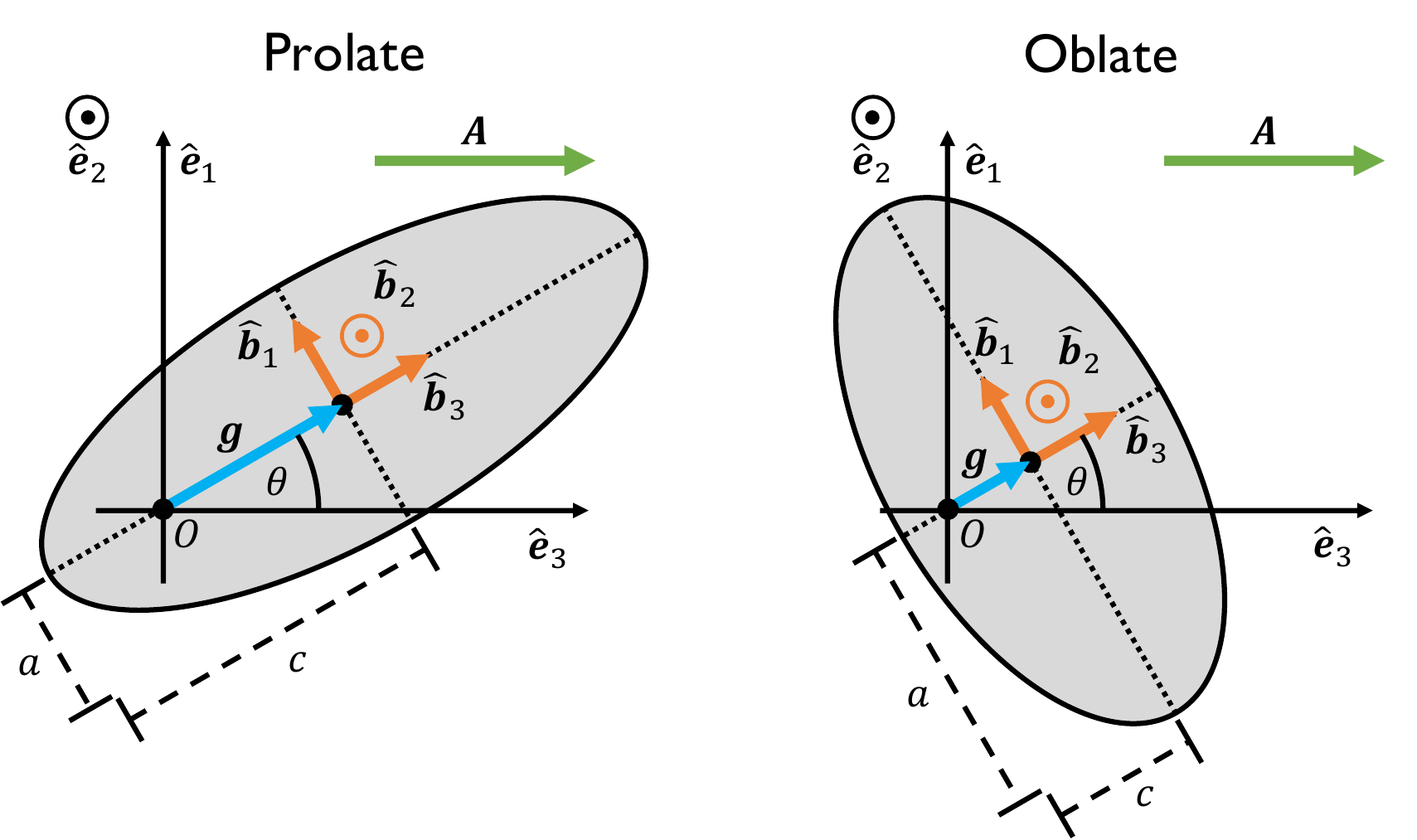}
    \caption{
        Schematic of a prolate and an oblate where the center of mass is shifted along the axis of symmetry. We use ellipses to represent the cross-sections of the two spheroids in the $e_{1}e_{3}$-plane. The quantity $a$ is the length from the spheroid center along the axis perpendicular to the axis of symmetry, while $c$ is the length from the center along the axis of symmetry. For the prolate, the axis of symmetry is the long axis ($c > a$), while for the oblate, the axis of symmetry is the short axis ($c < a$). $\vect{\hat{e}}_{1}$, $\vect{\hat{e}}_{2}$, and $\vect{\hat{e}}_{3}$ are lab frame unit vectors where $\vect{\hat{e}}_{2}$ points out of the page. $\vect{\hat{b}}_{1}$, $\vect{\hat{b}}_{2}$, and $\vect{\hat{b}}_{3}$ are body frame unit vectors, where $\vect{\hat{b}}_{2}$ also points out of the page and $\vect{\hat{b}}_{3}$ is along the axis of symmetry. The offset of the center of the spheroid from the center of mass $O$ is $\vect{g}$. Since the center of mass is shifted along the axis of symmetry, $\vect{g} \parallel \vect{\hat{b}}_{3}$. The green arrow is the direction of the gas flow $\vect{A}$. 
    }
    \label{fig:spheroid_schematic}
\end{figure*}

Thus far, we have not utilized any assumption about the geometry of the spheroid. 
We now evaluate each term in Eq.~\ref{eq:torque_three_terms_with_tensor} using a dedicated coordinate system shown in Fig.~\ref{fig:spheroid_schematic}.
The setup is equivalent to Sec.~\ref{sec:doublesphere}. 
Let $\vect{\hat{e}}_{1}$, $\vect{\hat{e}}_{2}$, and $\vect{\hat{e}}_{3}$ represent the Cartesian, unit directions of the lab frame that is centered at the origin $O$. 
We define $\vect{\hat{e}}_{3}$ to be along the direction of gas flow $\vect{A}$ and let the grain rotate around $\vect{\hat{e}}_{2}$. 

Let the body frame of the spheroid also be a Cartesian coordinate system with its center denoted by $G$. We use $\vect{\hat{b}}_{1}$, $\vect{\hat{b}}_{2}$, and $\vect{\hat{b}}_{3}$ to denote the unit vectors. $\vect{\hat{b}}_{3}$ is along the axis of asymmetry and $\vect{\hat{b}}_{2}$ is parallel to $\vect{\hat{e}}_{2}$. The angle between $\vect{\hat{b}}_{3}$ and $\vect{\hat{e}}_{3}$ is $\theta$. The relation between these two sets of unit vectors is simply
\begin{align}
    \vect{\hat{b}}_{i} = W_{ij} \vect{\hat{e}}_{j}
\end{align}
where we adopt the Einstein summation convention and $i$ and $j$ are indices from 1 to 3. $W_{ij}$ are the elements of the matrix
\begin{align} \label{eq:W_unitdir_convert}
    \matr{W} = 
    \begin{pmatrix}
        \cos \theta & 0 & - \sin \theta \\
        0 & 1 & 0 \\
        \sin \theta & 0 & \cos \theta
    \end{pmatrix}. 
\end{align}

In the body frame, the surface of the spheroid follows 
\begin{align} \label{eq:spheroid_surface}
    \frac{ b_{1}^{2} + b_{2}^{2} }{ a^{2} } + \frac{ b_{3}^{2} }{ c^{2} } = 1
\end{align}
where $b_{1}$, $b_{2}$, and $b_{3}$ are coordinates in the body frame along $\vect{\hat{b}}_{1}$, $\vect{\hat{b}}_{2}$, and $\vect{\hat{b}}_{3}$, respectively. In addition, $c$ is the length of the spheroid along the axis of symmetry (not to be confused with the isothermal sound speed $c_{s}$ or the speed of light) and $a$ is the length along the axis perpendicular to that. A prolate has $c>a$, while an oblate has $c<a$. 
Note that to fulfill the assumption of drag in the Epstein regime, the longest axis of the spheroid should be smaller than the mean free path of the gas (which is estimated in Sec.~\ref{sec:disk_alignment_timescale} below for an illustrative disk).

Considering axisymmetric particles makes the surface integral easier to compute via a surface of revolution. We parameterize the surface by introducing the azimuthal angle $\phi$ around the $\vect{\hat{b}}_{3}$ axis
\begin{align} \label{eq:spheroid_surface_parameterized}
    \begin{pmatrix}
        b_{1} \\
        b_{2}
    \end{pmatrix}
    = 
    \begin{pmatrix}
        w \cos \phi \\
        w \sin \phi
    \end{pmatrix}
\end{align}
where $\phi$ starts from $\vect{\hat{b}}_{1}$ and increases following the right-hand rule. 
Furthermore, $w$ is the perpendicular distance from the surface to the axis of symmetry, which depends on $b_{3}$, and serves as the generatrix for the surface integration: 
\begin{align} \label{eq:spheroid_generatrix}
    w(b_{3}) \equiv a \sqrt{1 - \bar{b}_{3}^{2}}
\end{align}
where $\bar{b}_{3} \equiv b_{3} / c$ is $b_{3}$ normalized by $c$. 
We will use the bar symbol to denote length quantities normalized by $c$. With the generatrix, we can derive the surface element to be
\begin{align}
    d \sigma = a c \sqrt{ 1 + (\bar{a}^{2} - 1) \bar{b}_{3}^{2} } d \bar{b}_{3} d \phi
\end{align}
where $\bar{a} \equiv a / c$. 

The normal vector of the spheroid is related to the gradient of Eq.~\ref{eq:spheroid_surface}. We express the unit normal vector by $\vect{n} = \tilde{n}_{i} \vect{\hat{b}}_{i}$: 
\begin{align}
    \begin{pmatrix}
        \tilde{n}_{1} \\
        \tilde{n}_{2} \\
        \tilde{n}_{3} 
    \end{pmatrix}
    = 
    \frac{1}{\sqrt{1 + (\bar{a}^{2} - 1) \bar{b}_{3}^{2}}}
    \begin{pmatrix}
        \sqrt{1 - \bar{b}_{3}^{2}} \cos \phi \\
        \sqrt{1 - \bar{b}_{3}^{2}} \sin \phi \\
        \bar{a} \bar{b}_{3}
    \end{pmatrix}
    \text{ .}
\end{align}
We use a tilde to denote components using $\vect{\hat{b}}_{i}$ as basis vectors.

The radius vector of the surface from $O$ can be decomposed to
\begin{align} \label{eq:spheroid_surface_radius_vector}
    \vect{r} = \vect{g} + \vect{s}
\end{align}
where $\vect{g}$ points from the center of mass to the geometric center $G$ and $\vect{s}$ points from $G$ to the surface. The components of $\vect{s}$ using $\vect{\hat{b}}_{i}$ basis vectors are 
\begin{align}
    \begin{pmatrix}
        \tilde{s}_{1} \\
        \tilde{s}_{2} \\
        \tilde{s}_{3}
    \end{pmatrix}
    = 
    \begin{pmatrix}
        a \sqrt{1 - \bar{b}_{3}^{2}} \cos \phi \\
        a \sqrt{1 - \bar{b}_{3}^{2}} \sin \phi \\
        c \bar{b}_{3}
    \end{pmatrix}
\end{align}
For the offset vector $\vect{g}$, we first consider the offset along the axis of symmetry to follow the strict assumption of axisymmetry. If the length between $O$ and $G$ is $g$, we have $\vect{g} = g \vect{\hat{b}}_{3}$. We later relax this assumption and explore the effects of $\vect{g}$ when it is not along the axis of symmetry. 

With the relevant details of the surface defined, we can now solve for $\vect{K}$, $\matr{L}$, and $\matr{M}$. 
When integrating over the surface, we find a common occurrence of an integral in the form of
\begin{align} \label{eq:integral_E}
    E[f(x)] \equiv \int_{-1}^{1} \frac{f(x)}{ \sqrt{1 + (\bar{a}^{2} - 1) x^{2}} } d x
\end{align} 
where $x$ is a dummy variable for integration and $f(x)$ is some function that depends on $x$. 
The dummy variable $x$ originates from $\bar{b}_{3}$ as we integrate over the surface. 
The definition of $E[f(x)]$ implicitly depends on $\bar{a}$ for easier notation. 
Appendix~\ref{sec:E_solutions} shows the analytical solutions to the $E$-integrals used in this paper. 

For $\vect{K}$, we make use of Eq.~\ref{eq:spheroid_surface_radius_vector} for $\vect{r}$ and obtain: 
\begin{align}
    \vect{K} 
    = \vect{g} \times \oint_{S} \vect{n} d \sigma + \oint_{S} \vect{s} \times \vect{n} d \sigma 
    = 0. 
\end{align}
Note that the integral in the first term, $\oint_{S} \vect{n} d \sigma$, is always $\vect{0}$ for a closed surface\footnote{Under the current context, this is true since we assumed that the entire surface of a grain reflects gas equally through Eq.~\ref{eq:force_per_unit_area}. If certain patches of the grain reflect gas differently, then there is an interesting possibility that the integral may not have to be $\vect{0}$. }. The latter term integrates to 0 for a spheroid. 

For $\matr{L}$, using Eq.~\ref{eq:spheroid_surface_radius_vector} gives four terms: 
\begin{align} \label{eq:spheroid_matr_L}
    \matr{L}
    = \vect{g} \times \oint_{S} \matr{N} d \sigma (\vect{g} \times) 
    + \vect{g} \times \oint_{S} \matr{N} (\vect{s} \times) d \sigma 
    + \oint_{S} \vect{s} \times \matr{N} d \sigma (\vect{g} \times) \nonumber \\
    + \oint_{S} \vect{s} \times \matr{N} (\vect{s} \times) d \sigma. 
\end{align}
It is convenient to know that the integral of $\matr{N}$ expressed in $\vect{\hat{b}}_{i}$ basis vectors is 
\begin{align} \label{eq:oint_N_dsigma}
    \oint_{S} \matr{N} d \sigma = 
    \pi c^{2}
    \begin{pmatrix}
        \bar{a} E[1 - x^{2}] & 0 & 0 \\
        0 & \bar{a} E[1 - x^{2}] & 0 \\
        0 & 0 & 2 \bar{a}^{3} E[x^{2}]
    \end{pmatrix}. 
\end{align}
For the first term of Eq.~\ref{eq:spheroid_matr_L}, using the $\vect{\hat{b}}_{i}$ basis vectors, we find that
\begin{align}
    \vect{g} \times \oint_{S} \matr{N} d \sigma (\vect{g} \times) 
    = - \pi c^{4} \bar{a} \bar{g}^{2} E[1 - x^{2}]
    \begin{pmatrix}
        1 & 0 & 0 \\
        0 & 1 & 0 \\
        0 & 0 & 0 
    \end{pmatrix}
\end{align}
where $\bar{g} \equiv g / c$. 
Each element in the second term integrates to $0$. Since $\matr{N}$ is a symmetric 2-rank tensor and both $\vect{s} \times$ and $\vect{g} \times$ are anti-symmetric 2-rank tensors, the third term is equivalent to the transpose of the second term. 
Lastly, the fourth term is a symmetric tensor, which means there are only six independent elements when expressed as a matrix in general. 
For the spheroid, four of the independent elements integrate to 0 leaving only two non-zero elements: 
\begin{align}
    \oint_{S} \vect{s} \times \matr{N} (\vect{s} \times) d \sigma 
    = - \pi c^{4} \bar{a} (1 - \bar{a}^{2})^{2} E[x^{2} - x^{4}]
    \begin{pmatrix}
        1 & 0 & 0 \\
        0 & 1 & 0 \\
        0 & 0 & 0 
    \end{pmatrix}
\end{align}
Thus, we find that 
\begin{align} \label{eq:spheroid_L}
    \tilde{L}_{11} = \tilde{L}_{22}
    = - \pi c^{4} \bar{a} \bigg[ \bar{g}^{2} E[1 - x^{2}] + (1 - \bar{a}^{2})^{2} E[x^{2} - x^{4}] \bigg]
\end{align}
while the rest of the elements are zero. 

Lastly, for $\matr{M}$, using the $\vect{\hat{b}}_{i}$ basis vectors, we have 
\begin{align} \label{eq:spheroid_M}
    \matr{M} 
    &= \vect{g} \times \oint_{S} \matr{N} d \sigma + \oint_{S} \vect{s} \times \matr{N} d \sigma \nonumber \\
    &= \pi c^{3} \bar{a} \bar{g} E[1 - x^{2}]
    \begin{pmatrix}
        0 & -1 & 0 \\
        1 & 0 & 0 \\
        0 & 0 & 0 
    \end{pmatrix}
\end{align}
where we found that the second term is $\matr{0}$. 

We can now derive the torque from Eq.~\ref{eq:torque_three_terms_with_tensor}. 
Given the assumed $\vect{A} = A \vect{\hat{e}}_{3}$, we can express $\vect{A}=\tilde{A}_{i} \vect{\hat{b}}_{i}$ and find that 
\begin{align}
    \vect{A} = - A \sin \theta \vect{\hat{b}}_{1} + A \cos \theta \vect{\hat{b}}_{3}
\end{align}
using Eq.~\ref{eq:W_unitdir_convert}. Since we only consider rotation around $\vect{\hat{e}}_{2}$ and $\vect{\hat{e}}_{2}=\vect{\hat{b}}_{2}$, the angular velocity is simply $\vect{\omega} = \omega \vect{\hat{e}}_{3} = \omega \vect{\hat{b}}_{2}$. 
Applying to Eq.~\ref{eq:torque_three_terms_with_tensor}, the torque is thus
\begin{align}
    \tilde{\Gamma}_{2} = 
    &- \rho_{g} \vth \pi c^{4} \omega \bar{a} 
        \bigg[ \bar{g}^{2} E[1 - x^{2}] + (1 - \bar{a}^{2})^{2} E[x^{2} - x^{4}] \bigg] \nonumber \\
    &- \rho_{g} \vth A \pi c^{3} \bar{a} \bar{g} E[1 - x^{2}] \sin \theta 
\end{align}
and we find that $\tilde{\Gamma}_{1} = \tilde{\Gamma}_{3} = 0$. 
Note that $\Gamma_{2} = \tilde{\Gamma}_{2}$ also because $\vect{\hat{e}}_{2} = \vect{\hat{b}}_{2}$. 
Thus, we see that the torque is only limited along $\vect{\hat{e}}_{2}$ which is what we expect when we only consider $\vect{\omega}$ along $\vect{\hat{e}}_{2}$.

The moment of inertia around the axis passing through the geometric center $G$ and perpendicular to the axis of symmetry is 
\begin{align} \label{eq:I_perpendicular}
    I_{\perp} = \frac{4 \pi}{15} c^{5} \rho_{s} \bar{a}^{2} (1 + \bar{a}^{2}) 
\end{align}
for a homogenous spheroid with material density $\rho_{s}$. 
However, to shift the center of mass $O$ away from the geometric center $G$, the spheroid cannot be homogenous. 
Nevertheless, based on the double-sphere model, the necessary shift is relatively small compared to the size of the grain, so the homogenous sphere remains a good approximation even in this case.
Using $I_{\perp}$ and the parallel axis theorem, we have
\begin{align} \label{eq:I_spheroid_with_g}
    I_{s} = \frac{4 \pi}{15} c^{5} \rho_{s} \bar{a}^{2} (1 + \bar{a}^{2} + 5 \bar{g}^{2} ) 
\end{align}
where we can see that $I_{s}$ corresponds to $I_{\perp}$ when $\bar{g} \ll 1$. 
As an approximation, we simply adopt $I_{s}$ as the moment of inertia for the spheroid.
The subscript ``s'' stands for ``spheroid.''

The equation of motion has the same form as the double-sphere model:
\begin{align} \label{eq:equation_of_motion_spheroid}
    I_{s} \dot{\omega} + D_{s} \omega + P_{s} \sin \theta = 0 
\end{align}
where 
\begin{align}
    D_{s} &= \rho_{g} \vth \pi c^{4} \bar{a} \bigg[ \bar{g}^{2} E[1 - x^{2}] + (1 - \bar{a}^{2})^{2} E[x^{2} - x^{4}] \bigg] \nonumber \\
    P_{s} &= \rho_{g} \vth A \pi c^{3} \bar{a} \bar{g} E[1 - x^{2}] \nonumber
\end{align}
meaning the spheroid will also behave as a damped oscillator. 

We find that the potential energy is $U_{s} = - P_{s} \cos \theta$, which is again proportional to $- \cos \theta$ and means the final alignment angle is at $\theta=0$ for $P_{s}>0$ or at $\theta=\pi$ for $P_{s}<0$. Since $E[1 - x^{2}]$ is always positive for all $\bar{a} > 0$, the sign of $P_{s}$ comes entirely from the sign of $g$ and is independent of $\bar{a}$. In other words, both prolate and oblate grains will become aligned with their axis of symmetry along the gas flow, such that the direction from the center of mass to the geometric center follows the direction of the gas flow ($\vect{g} \parallel \vect{A}$). 

\begin{figure}
    \centering
    \includegraphics[width=\columnwidth]{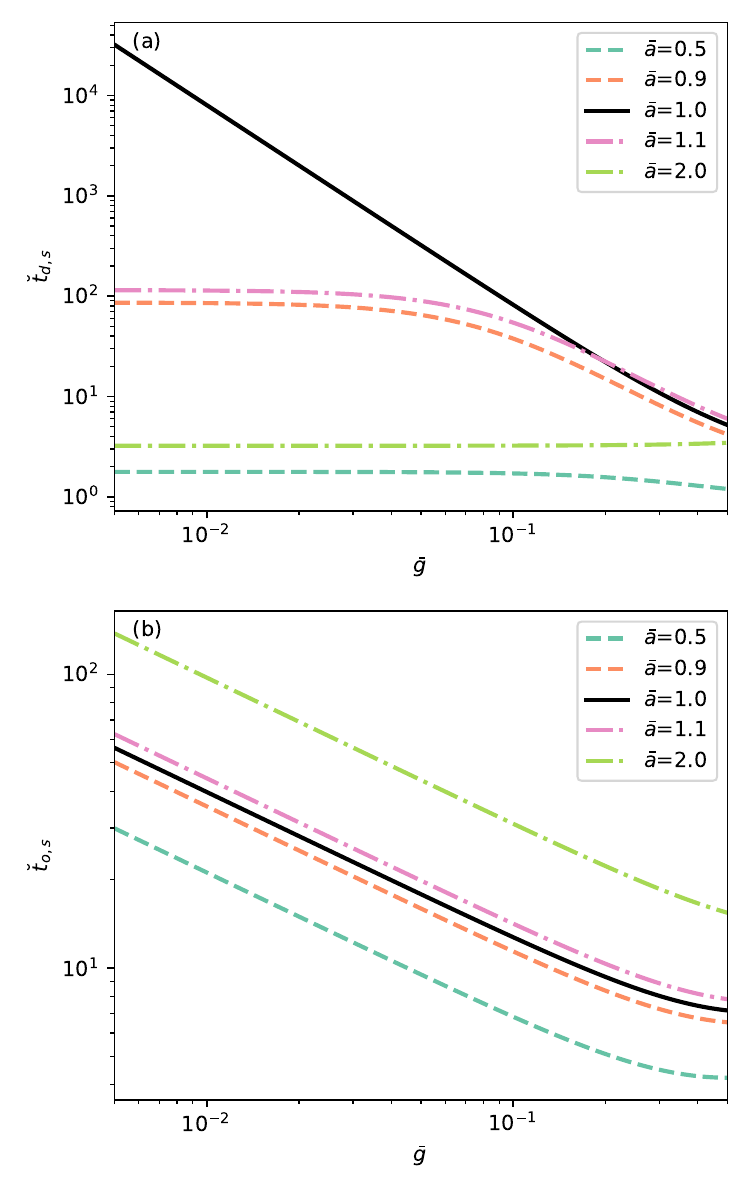}
    \caption{
        Dimensionless factors for the strictly axisymmetric spheroid as a function of $\bar{g}$ (the normalized distance from the center of mass to the spheroid center, $g/c$) and $\bar{a}$ (the aspect ratio, $a/c$). Panel a: the damping time dimensionless factor, $\Breve{t}_{d,d}$. Panel b: the oscillation time dimensionless factor, $\Breve{t}_{o,d}$. Black solid lines correspond to the spherical case, $\bar{a}=1$. Dashed lines correspond to prolates ($\bar{a}<1$), while dash-dotted lines correspond to oblates ($\bar{a}>1$). 
    }
    \label{fig:spheroid_time_factor}
\end{figure}

Given Eq.~\ref{eq:equation_of_motion_spheroid} we can explore the damping time $t_{d}$ and oscillation time $t_{o}$ analytically like in the previous section. 
We find that 
\begin{align} \label{eq:spheroid_t_d}
    t_{d, s} &= \frac{ \rho_{s} c}{ \rho_{g} \vth } \Breve{t}_{d,s} , \nonumber \\ 
    \Breve{t}_{d,s} &\equiv
        \frac{8 }{15} \frac{ \bar{a} (1 + \bar{a}^{2} + 5 \bar{g}^{2}) }{ \bar{g}^{2} E[1 - x^{2}] + (1 - \bar{a}^{2})^{2} E[x^{2} - x^{4}] }
    . 
\end{align}
The damping time $t_{d,s}$ is once again related to the stopping time modified by a dimensionless quantity that depends on the structure of the grain $\Breve{t}_{d,s}$ (through $\bar{a}$ and $\bar{g}$). 
Fig.~\ref{fig:spheroid_time_factor}a shows how $\Breve{t}_{d,s}$ depends on $\bar{a}$ and $\bar{g}$. 
If we have a sphere ($\bar{a}=1$) rotating about its geometric center ($\bar{g}=0$), $\Breve{t}_{d,s} \rightarrow \infty$ (or equivalently, $D_{s} \rightarrow 0$) meaning it will not be damped by the gas at all. 
The result makes sense since the normal direction at each point of the sphere is always directed radially from the center of the sphere and thus the gas cannot produce any torque as the sphere spins. 
However, if the center of mass of the sphere does not correspond to the geometrical center ($\bar{g} \neq 0$), then $t_{d,s}$ is no longer infinite and its rotation can be damped. 
If we have a homogenous spheroid ($\bar{a} \neq 1$) rotating about its geometric center ($\bar{g}=0$), $\Breve{t}_{d,s}$ is also finite since the normal directions of most of the surface elements are no longer strictly in the radial direction from $O$ and gas can produce a torque. 
In general, as the non-sphericity ($|\bar{a}-1|$) increases, $\Breve{t}_{d,s}$ decreases. 
In other words, more non-spherical grains (for both prolates and oblates) are better damped by the gas and are aligned faster. 
With $\bar{g}=0.01$ and $\bar{a}=0.9$, $\Breve{t}_{d,s} \sim 80$. 

The oscillation time is 
\begin{align} \label{eq:spheroid_t_o}
    t_{o,s} &= \sqrt{ \frac{\rho_{s} c^{2}}{\rho_{g} \vth A} } \Breve{t}_{o,s} \nonumber \\ 
    \Breve{t}_{o,s} &\equiv \sqrt{ \frac{16 \pi^{2} }{15} \frac{ \bar{a} (1 + \bar{a}^{2} + 5 \bar{g}^{2}) }{\bar{g} E[1 - x^{2}]} }
\end{align}
which is also similar to the expression for the double-sphere. 
Fig.~\ref{fig:spheroid_time_factor}b shows how $\Breve{t}_{o,s}$ depends on the structure. 
For a given $\bar{a}$, increasing $\bar{g}$ decreases $t_{o,s}$ for both prolates and oblates, which means the key to allowing a spheroid to oscillate lies in the offset between the center of mass from the geometric center. 
For $\bar{g}\rightarrow0$, $\Breve{t}_{o,s} \rightarrow \infty$ (or equivalently, $P_{s} \rightarrow 0$) meaning the grain will not oscillate without the offset. 
Fig.~\ref{fig:spheroid_time_factor}b also shows that increasing $\bar{a}$ (more oblong) increases $\Breve{t}_{o,d}$. However, the large difference is in part due to keeping the length along the axis of symmetry, $c$, fixed (long axis for a prolate, but short axis for an oblate). With $\bar{g}=0.01$ and $\bar{a}=0.9$, $\Breve{t}_{o,s} \sim 35$. 

\subsection{Quasi-Axisymmetric Spheroid} \label{sec:quasi_axisymmetric_spheroid}

Previously, we have enforced strict axisymmetry which requires the center of mass to be along the axis of symmetry. 
As a result, both prolate and oblate cases will be aligned with their axis of symmetry along the gas flow. 
Since polarization observations of disks favor effectively prolate grains, we would naively continue our discussion with prolate grains. 
However, if oblate grains are aligned along some long axis and the direction of the short axis is random as an ensemble, they can also appear effectively prolate around the azimuth of a disk if the dust-gas drift velocity is azimuthal. 
We are particularly interested in verifying if oblate grains can be aligned to the gas drift along the long axis if $\vect{g}$ is along the long axis.

The benefits of axisymmetry lie in the integration of the surface and $\vect{g}$ itself does not participate in the integration. 
In other words, we can allow the center of mass to be anywhere in the spheroid, while keeping most of the derivation the same. 
If $\vect{g}$ is not along the axis of symmetry, the grain is no longer strictly axisymmetric in terms of the distribution of mass. However, since the surface remains axisymmetric, we describe the model as a ``quasi-axisymmetric'' model. 
We show that both the shape parameter $\bar{a}$ and the offset vector $\vect{g}$ determine the final alignment direction. 
In the case where $\vect{g}$ lies along the long axis of an oblate, we find that the long axis of the oblate will be aligned with the gas flow.

\begin{figure}
    \centering
    \includegraphics[width=0.8\columnwidth]{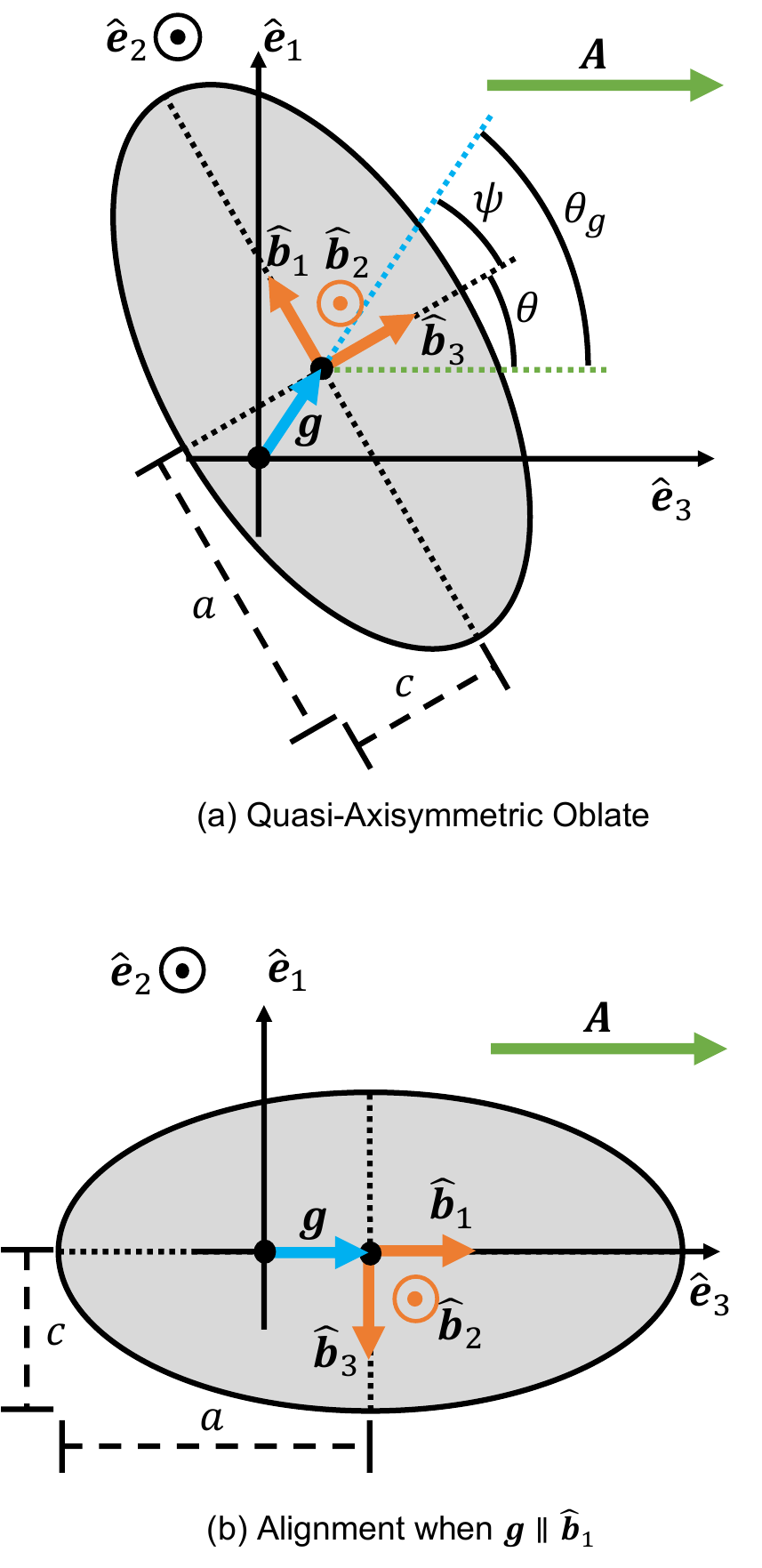}
    \caption{
        Panel a: a schematic of the coordinate system used to describe the quasi-axisymmetric oblate following the same coloring scheme and notation as Fig.~\ref{fig:spheroid_schematic}. Without strictly requiring $\vect{g}$ to be parallel to $\vect{\hat{b}}_{3}$, it can form an angle $\psi$ from $\vect{\hat{b}}_{3}$. The green dotted line helps visualize the angle $\theta$ formed by $\vect{\hat{b}}_{3}$ and $\vect{\hat{e}}_{3}$. The blue dotted line is in the direction of $\vect{g}$ which helps visualize $\psi$. 
        Panel b: a schematic of a quasi-axisymmetric oblate with $\vect{g} \parallel \vect{\hat{b}}_{1}$ (or equivalently $\psi=90^{\circ}$) when reaching alignment at $\theta=-90^{\circ}$.  
    }
    \label{fig:quasiaxisymmetric_oblate}
\end{figure}

Following the same notation above, consider $\vect{g}$ that forms an angle $\psi$ from $\vect{\hat{b}}_{3}$. Fig.~\ref{fig:quasiaxisymmetric_oblate}a shows a schematic of a quasi-axisymmetric oblate. We pick $\vect{\hat{b}}_{1}$ to be in the same plane as $\vect{g}$ and $\vect{\hat{b}}_{3}$, which gives
\begin{align} \label{eq:offset_vector_with_alpha}
    \begin{pmatrix}
        \tilde{g}_{1} \\
        \tilde{g}_{2} \\
        \tilde{g}_{3}
    \end{pmatrix}
    = 
    \begin{pmatrix}
        g \sin \psi \\
        0 \\
        g \cos \psi
    \end{pmatrix}
\end{align}
using $\vect{\hat{b}}_{i}$ as basis vectors. 
The gas flow remains in the $\vect{\hat{e}}_{3}$ direction. We can choose $\vect{\hat{e}}_{1}$ such that $\vect{g}$ lies in the plane formed by $\vect{\hat{e}}_{1}$ and $\vect{\hat{e}}_{3}$. $\vect{\hat{e}}_{2}$ remains parallel to $\vect{\hat{b}}_{2}$.

We first evaluate $\vect{K}$, $\matr{L}$, and $\matr{M}$ using Eq.~\ref{eq:offset_vector_with_alpha}. 
We find that $\vect{K} = \vect{0}$. For $\matr{L}$, one only needs to re-evaluate the first term that depends on $\vect{g}$ and is non-zero. We find that
\begin{align}
    \tilde{L}_{11} &= - \pi c^{4} \bar{a} \bar{g}^{2} E[1 - x^{2}] \cos^{2} \psi \nonumber \\
    \tilde{L}_{13} = \tilde{L}_{31} &= - \pi c^{4} \bar{a} \bar{g}^{2} E[1 - x^{2}] \cos \psi \sin \psi \nonumber \\
    \tilde{L}_{22} &= - \pi c^{4} \bar{a} \bar{g}^{2}
        \bigg[ 
            E[1 - x^{2}] \cos^{2} \psi 
            + 2 \bar{a}^{2} E[x^{2}] \sin^{2} \psi 
        \bigg] \nonumber \\
    \tilde{L}_{33} &= - \pi c^{4} \bar{a} \bar{g}^{2} E[1 - x^{2}] \sin^{2} \psi 
\end{align}
while the rest of the elements of $\matr{L}$ are 0. When $\psi=0$, we recover Eq.~\ref{eq:spheroid_L}.
For $\matr{M}$, we find that 
\begin{align}
    \tilde{M}_{12} &= - \pi c^{3} \bar{a} \bar{g} E[1 - x^{2}] \cos \psi \nonumber \\
    \tilde{M}_{21} &= \pi c^{3} \bar{a} \bar{g} E[1 - x^{2}] \cos \psi \nonumber \\
    \tilde{M}_{23} &= - 2 \pi c^{3} \bar{a}^{3} \bar{g} E[x^{2}] \sin \psi \nonumber \\
    \tilde{M}_{32} &= \pi c^{3} \bar{a} \bar{g} E[1 - x^{2}] \sin \psi
\end{align}
while the rest of the elements of $\matr{M}$ are $0$. 
When $\psi=0$, we recover Eq.~\ref{eq:spheroid_M}. Lastly, the moment of inertia remains equal to Eq.~\ref{eq:I_spheroid_with_g}. 

\begin{figure}
    \centering
    \includegraphics[width=\columnwidth]{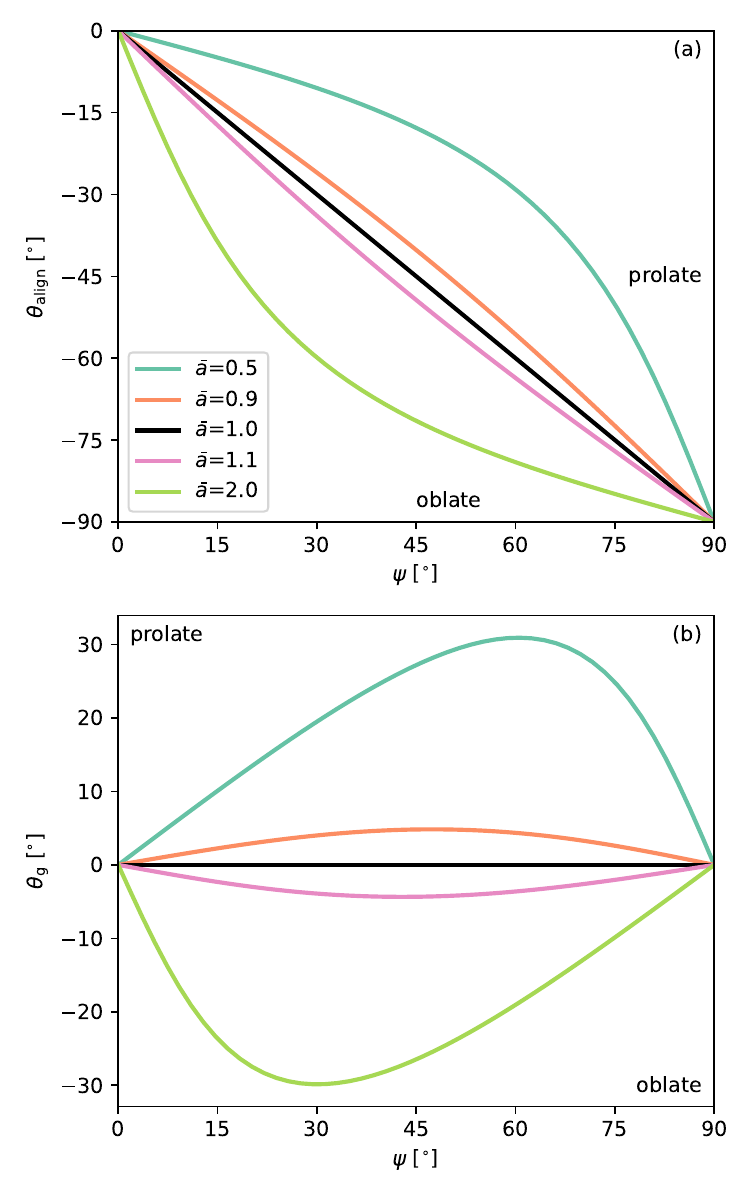}
    \caption{
        Panel a: the angle of alignment $\theta_{\text{align}}$ as a function of $\psi$ and for different values of $\bar{a}$ (a prolate means $\bar{a}<1$, while an oblate means $\bar{a}>1$).
        Panel b: the angle from $\vect{g}$ when at alignment. 
    }
    \label{fig:theta_align}
\end{figure}

The equation of motion becomes
\begin{align} \label{eq:quasi-axisymmetric_equation_of_motion}
    I_{s} \dot{\omega} + D_{q} \omega + P_{q} \sin \theta + P_{c} \cos \theta = 0
\end{align}
where the coefficients are 
\begin{align}
    D_{q} &= \rho_{g} \vth \pi c^{4} \bar{a} 
        \bigg[ 
            (1 - \bar{a}^{2})^{2} E[x^{2} - x^{4}]
            &+ \bar{g}^{2} E[1-x^{2}] \cos^{2} \psi \nonumber  \\
            & &+ 2 \bar{a}^{2} \bar{g}^{2} E[x^{2}] \sin^{2} \psi 
        \bigg]  \nonumber \\
    P_{q} &= \rho_{g} \vth A \pi c^{3} \bar{a} \bar{g} E[1-x^{2}] \cos \psi \nonumber \\
    P_{c} &= 2 \rho_{g} \vth A \pi c^{3} \bar{a}^{3} \bar{g} E[x^{2}] \sin \psi \text{ .} \nonumber 
\end{align}
The subscript ``q'' refers to the quasi-axisymmetric model, while the subscript ``c'' for $P_{c}$ refers to the coefficient of its cosine term (``$P_{q,c}$'' would be too redundant).
The existence of the extra $\cos \theta$ term in Eq.~\ref{eq:quasi-axisymmetric_equation_of_motion} alters the dynamical behavior giving a potential energy of
\begin{align}
    U_{q} = - P_{q} \cos \theta + P_{c} \sin \theta. 
\end{align}
We can derive the alignment angle by finding $\theta_{\text{align}}$ such that $U$ is a minimum giving
\begin{align} \label{eq:quasi_spheroid_alignment_angle}
    \theta_{\text{align}} 
    &= \atantwo \bigg( \frac{-P_{c}}{P_{q}} \bigg)  \nonumber \\
    &=  \atantwo \bigg( \frac{- 2 \bar{a}^{2} E[x^{2}] \sin \psi}{E[1 - x^{2}] \cos \psi} \bigg)
\end{align}
where $\text{atan2}$ is the arctangent that gives an angle in the correct quadrant by taking into account the sign of the arguments. 
The alignment angle does not depend on the length of $\vect{g}$, but only on $\psi$ and $\bar{a}$. Fig.~\ref{fig:theta_align}a shows the alignment angle as a function of $\psi$ for different values of $\bar{a}$. 
If $\psi=0$, i.e., $\vect{g}$ follows the axis of symmetry, then $P_{c}=0$ and $\theta_{\text{align}} = 0$ which recovers the previous results. 
If $\psi=\pi/2$, i.e., $\vect{g}$ follows the axis perpendicular to the axis of symmetry for both prolates and oblates, then $P_{q}=0$ and $\theta_{\text{align}} = - 90^{\circ}$. 
This means that an oblate will be aligned to $\vect{A}$ along its long axis if its center of mass is shifted along the long axis. 
A schematic of this conceptually important case is shown in Fig.~\ref{fig:quasiaxisymmetric_oblate}b for an oblate. 
In fact, if the center of mass is preferentially shifted toward the long axis of the oblate ($\psi > \sim 45^{\circ}$), then the grain is preferentially aligned along the long axis which is permitted by disk polarization when considering the ensemble. 
Eq.~\ref{eq:quasi_spheroid_alignment_angle} also suggests a prolate will be aligned to $\vect{A}$ along its short axis if $\vect{g}$ is shifted along that direction. 
We discuss the implications in Sec.~\ref{sec:discussion}. 


We can also derive the angle, $\theta_{g}$, that $\vect{g}$ forms with $\vect{A}$ which is shown in Fig.~\ref{fig:theta_align}b. 
Evidently, $\theta_{g}$ does not entirely follow $\vect{A}$, but is modified by the shape of the grain. 
Nevertheless, it is within $\sim 30^{\circ}$ for the range of $\bar{a}$ explored here.

\section{Alignment in a Disk} \label{sec:disk}

Sec.~\ref{sec:doublesphere} and \ref{sec:spheroid} have established how grains are aligned and how the alignment timescales depend on environmental factors. 
This section will now implement the axisymmetric spheroid model in a simple disk environment to assess the alignment timescales. 
Observations of polarization permit either prolate grains aligned with the long axis toroidally around the disk or oblate grains also aligned with the long axis toroidally around the disk. 
For brevity, we will only apply the results from the strictly axisymmetric prolate case. 
We first describe how the gas and dust velocities depend on the properties of the disk and grains in Sec.~\ref{sec:disk_alignment_St_alpha}. 
In Sec.~\ref{sec:disk_alignment_timescale}, we implement the velocity profiles and analyze the alignment timescales across the disk. 

\subsection{Disk Velocity} \label{sec:disk_alignment_St_alpha}

We assume that the dynamics of gas are not affected by the dust. 
We use ($R,\Phi$) to denote the cylindrical radius and azimuth of the disk. 
Let the Keplerian frequency in the midplane be $\Omega_{K} \equiv \sqrt{G M_{*} / R^{3}}$ where $M_{*}$ is the stellar mass. 
It is also convenient to define the Keplerian velocity $v_{K} \equiv \Omega_{K} R$. 
We define a pressure scale height as $H \equiv c_{s} / \Omega_{K}$ and let $P_{g}$ be the gas pressure. 
The azimuthal velocity of the gas in the midplane follows from the radial force balance: 
\begin{align} \label{eq:v_g_phi}
    v_{g, \Phi} = v_{K} \sqrt{1 - \eta}
\end{align}
where the dimensionless factor
\begin{align} \nonumber
    \eta 
    \equiv - \frac{1}{ \Omega_{K}^{2} \rho_{g} R} \frac{\partial P_{g} }{\partial R} 
    = - \bigg(\frac{H}{R} \bigg)^{2} \frac{\partial \ln P_{g} }{\partial \ln R} 
\end{align}
represents the level of deviation from $v_{K}$ due to the pressure gradient \citep{Nakagawa1986Icar...67..375N, Armitage2015arXiv150906382A}. 
It is convenient to define the pressure gradient as $\beta \equiv \partial \ln P_{g} / \partial \ln R$. 
We assume the gaseous disk is accreting and parameterize an inward flow by 
\begin{align} \label{eq:v_g_R}
    v_{g, R} = - \alpha \bigg(\frac{H}{R} \bigg)^{2} v_{K}
\end{align}
where $\alpha$ is a dimensionless parameter\footnote{Note that the $\alpha$ here is simply a parameterization for the inward flow since we do not presume the origin. Thus, it is not immediately related to the typical $\alpha$-viscosity prescription from \cite{Shakura1973A&A....24..337S}. Nevertheless, if viscosity serves as the origin of the inward flow, then our parameter $\alpha$ should only differ from the $\alpha$-viscosity by a factor of 1.5. }.

The azimuthal velocity of the dust is \citep{Armitage2015arXiv150906382A}:
\begin{align} \label{eq:v_d_phi}
    v_{d, \Phi} = v_{g, \Phi} - \frac{1}{2} \St v_{d, R}
\end{align}
where $v_{d, R}$ is the radial velocity of the dust and $\St$ is the Stokes number of the dust. 
The radial velocity of the dust is 
\begin{align} \label{eq:v_d_R}
    v_{d, R} = \frac{v_{g, R} - \St \eta v_{K} }{ \St^{2} + 1 }. 
\end{align}

With these expressions, we can derive the aerodynamic flow of gas experienced by the dust $\vect{A}$ through:
\begin{align}
    A_{R} &\equiv v_{g, R} - v_{d, R} = - \frac{\alpha \St + \beta}{\St + \St^{-1}} \bigg( \frac{H}{R} \bigg)^{2} v_{K} \\
    A_{\Phi} &\equiv v_{g, \Phi} - v_{d, \Phi} = \frac{1}{2} \frac{ - \alpha + \beta \St }{\St + \St^{-1}} \bigg( \frac{H}{R} \bigg)^{2} v_{K}
\end{align}
in the radial and azimuthal directions, respectively. The flow speed is simply $A = \sqrt{A_{R}^{2} + A_{\Phi}^{2}}$. We can also obtain the flow direction $\chi$ which determines the alignment direction of the grain:
\begin{align}
    \chi = \atantwo \bigg[ \frac{1}{2} \frac{ - \alpha + \beta \St }{ - (\alpha \St + \beta) } \bigg]
\end{align}
where $\chi=0$ means the dust feels an outward radial gas flow, while $\chi=90^{\circ}$ means the dust feels an azimuthally directed flow along the direction of disk rotation (i.e., a tailwind if the dust orbits azimuthally). 
Since prolate grains are aligned along the long axis, the polarization angle will be parallel to $\chi$, i.e., the polarization angle has a $180^{\circ}$ degeneracy and only ranges from $0^{\circ}$ to $180^{\circ}$. 

\begin{figure*}
    \centering
    \includegraphics[width=\textwidth]{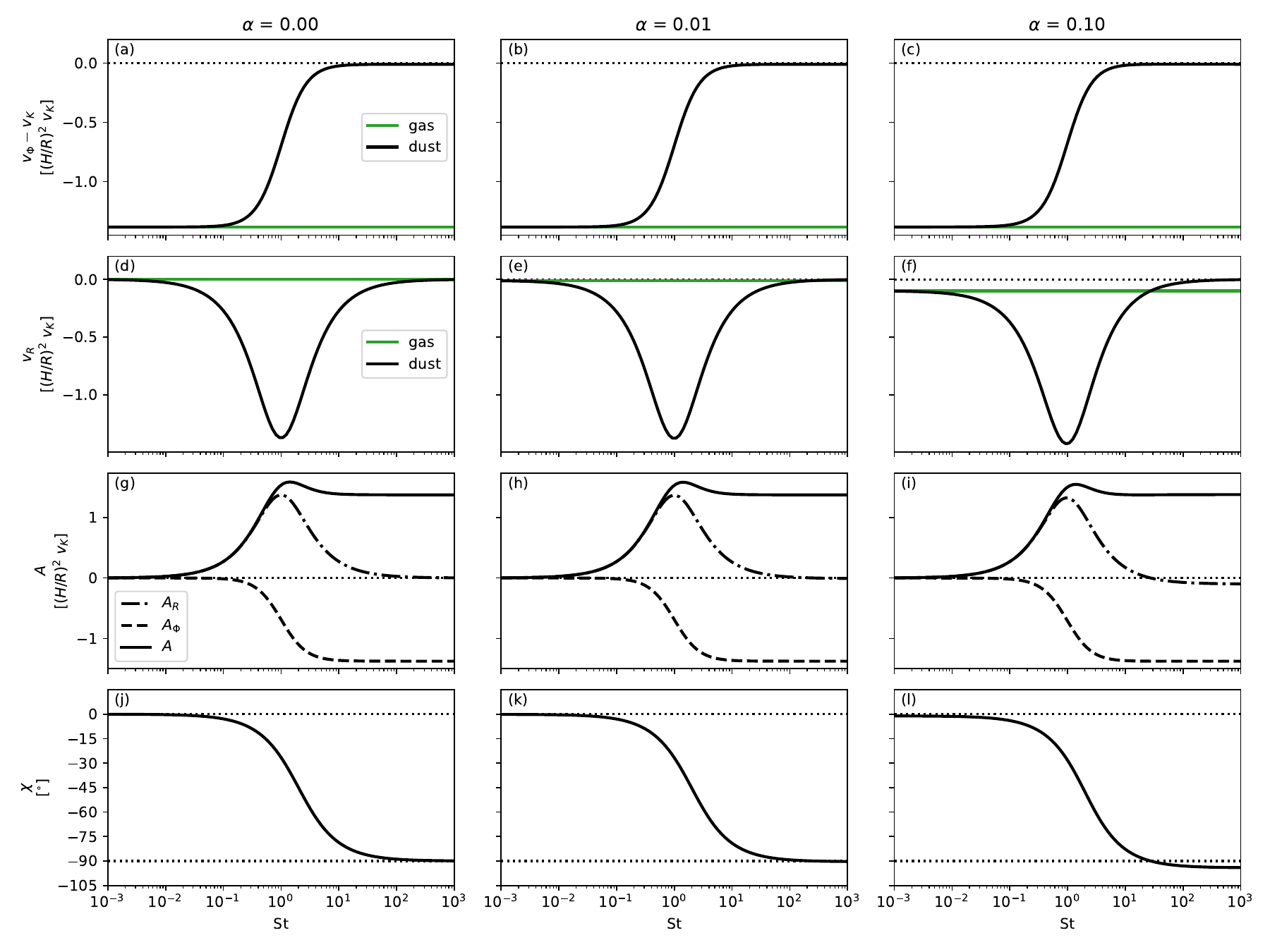}
    \caption{
        Radial and azimuthal velocity profiles as a function of $\St$ for different levels of gas accretion, parameterized by $\alpha$. The left to right columns correspond to $\alpha=0$ (no gas accretion), 0.01, and 0.1. The adopted pressure gradient, $\beta$, is -2.75.
        Top row: the azimuthal velocity subtracted by the Keplerian velocity $v_{K}$ normalized by $(H/R)^2 v_{K}$. The black and green solid lines are the profiles for the dust and gas, respectively. The horizontal dotted line corresponds to $v_{\Phi}$ with Keplerian rotation. 
        Second row: the radial velocity for the dust and gas (black and green lines, respectively). The horizontal dotted line corresponds to $v_{R}=0$. 
        Third row: the gas flow which is the gas velocity relative to the dust velocity. The radial and azimuthal components of the gas flow, $A_{R}$ and $A_{\Phi}$, are shown in solid and dashed lines, respectively. The horizontal dotted line corresponds to no relative velocity. 
        Fourth row: the gas flow direction $\chi$ in degrees. $\chi=0^{\circ}$ means radial, outward flow, while $\chi=90^{\circ}$ means azimuthal flow in the rotation direction. The horizontal dotted lines mark $0^{\circ}$ and $-90^{\circ}$ for visual guidance. 
    }
    \label{fig:St_alpha_grid}
\end{figure*}

Fig.\ref{fig:St_alpha_grid} shows the gas flow direction as a function of $\St$ for different levels of $\alpha$. 
When $\alpha=0$ (Fig.~\ref{fig:St_alpha_grid}, first column), the gas does not accrete ($v_{g,R}=0$), while $v_{g,\Phi}$ is always sub-Keplerian due to the pressure gradient $\beta$ from Eq.~\ref{eq:v_g_phi} (Fig.~\ref{fig:St_alpha_grid}a, d). 
The velocity of the dust depends on $\St$. When $\St \ll 1$, the dust is completely coupled to the gas and the dust velocity equals the gas velocity. 
The gas flow, $\vect{A}$, is thus close to $\vect{0}$ (Fig.~\ref{fig:St_alpha_grid}g). 
As $\St$ increases to $\sim 1$, $v_{d,\Phi}$ begins to increase (becomes less negative) and $v_{d,R}$ drops quickly. 
This is the well-known scenario where dust grains move inwards quickly when $\St \sim 1$ (i.e., the radial drift; \citealt{Armitage2015arXiv150906382A}). 
The gas flow is predominantly in the radial direction making $\chi \sim 0^{\circ}$ for relatively small particles with $\St \ll 1$ (Fig.~\ref{fig:St_alpha_grid}j). 
When $\St \gg 1$, the grains decouple from the gas and maintain a Keplerian orbit ($v_{d,\Phi}=v_{K}$, $v_{d,R}=0$). 
The gas flow becomes opposite to the azimuthal direction making $\chi \sim -90^{\circ}$ and the dust feels an azimuthal headwind (Fig.~\ref{fig:St_alpha_grid}j). 
Thus, for prolate grains aligned to the gas flow along the long axis, the polarization direction should be parallel to the radial direction when $\St \ll 1$ and parallel to the azimuthal direction when $\St \gg 1$. 

When $\alpha$ increases (Fig.~\ref{fig:St_alpha_grid}, second and third column), $v_{g,R}$ becomes non-zero and travels inward (Fig.~\ref{fig:St_alpha_grid}e, f).
However, $v_{g,\Phi}$ does not change, since it remains completely determined by the pressure gradient (Fig.~\ref{fig:St_alpha_grid}b, c). 
The main effect on the gas flow is when $\St \gg 1$ where the dust becomes Keplerian with $v_{d,R} \rightarrow 0$.
With a non-zero $v_{g,R}$, the dust eventually matches $v_{g,R}$ as the $\St$ increases (Fig.~\ref{fig:St_alpha_grid}e, f) 
and $v_{d,R}$ approaches zero from below, which leads to a characteristic Stokes number $\St_c$ in the high $\St$ regime where $A_R=0$ exactly:
\begin{align}
    \St_{c} 
    = - \frac{\beta}{ \alpha }
    .
\end{align}
The value of $\St_c$ is formally $\infty$ when $\alpha=0$ and of order $10^{2}$ when $\alpha=0.01$. 
It becomes $\sim 10$ when $\alpha$ increases to 0.1.
For grains with $\St>\St_c$, the alignment angle $\chi$ becomes more negative than $-90^\circ$, although the deviation remains relatively small, so that the alignment is approximately azimuthal.

The exploration shows that grains with large $\St$ can produce the azimuthal direction of polarization, and increasing $\alpha$ lessens the strict need for very large $\St$. 
At face value, this may serve as evidence of aerodynamically large grains that are fairly decoupled from the gas. 
However, even with $\alpha=0.1$, which would require a fairly large turbulence \citep{Rosotti2023NewAR..9601674R} or magnetic-wind driven accretion \citep[e.g.][]{Suriano2018MNRAS.477.1239S}, the required $\St$ is $\sim 20$ which is larger than what is typically considered.

The issue may depend on the assumed pressure gradient. Dust rings are fairly common \citep[e.g.][]{Andrews2018ApJ...869L..41A} and they can serve as traps for large grains \citep[e.g.][]{Pinilla2012A&A...538A.114P, Dullemond2018ApJ...869L..46D}. 
Gas kinematics also infer pressure bumps \citep{Teague2018ApJ...860L..12T}. 
With quickly varying pressure gradients along the radius across rings, $\beta$ may be closer to 0, for example, $\beta=0$ at pressure maxima and minima. 
Given how common dust rings are, we also explore the case when $\beta=0$ and show that small $\St$ can also produce gas flow parallel to the azimuthal direction. 

\begin{figure}
    \centering
    \includegraphics[width=\columnwidth]{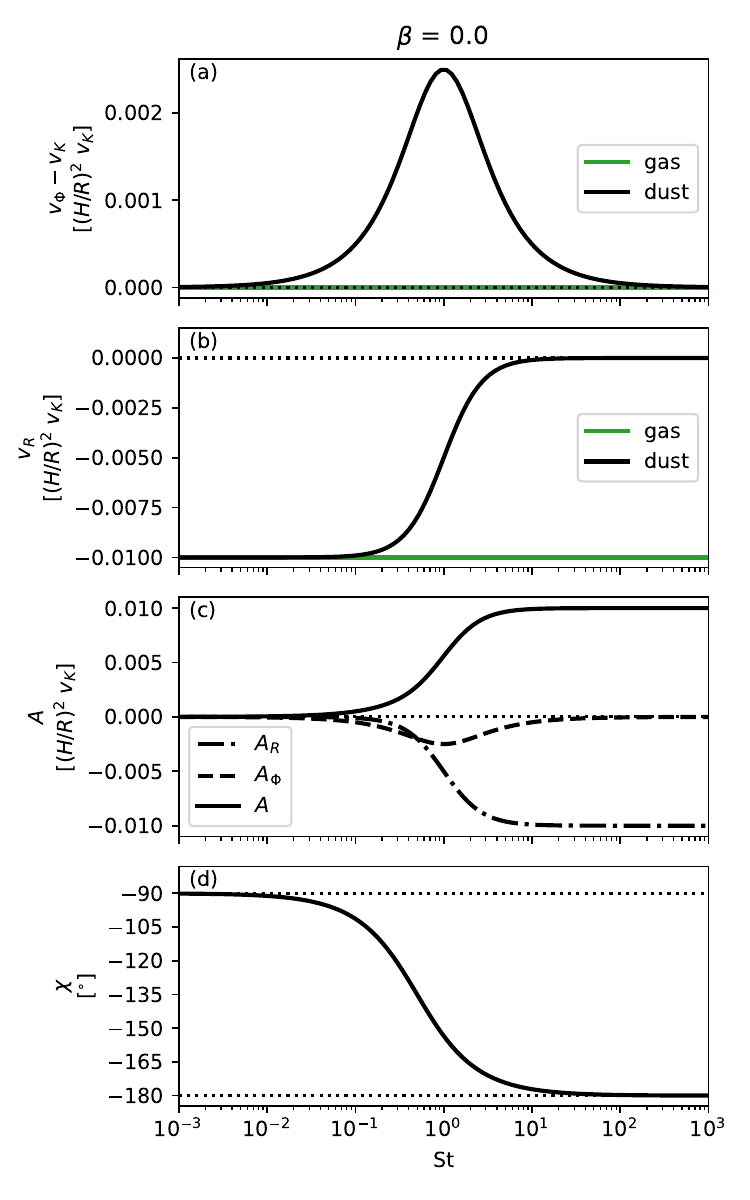}
    \caption{
        The velocity profiles when $\beta=0$ and $\alpha=0.01$ plotted in the same way as Fig.~\ref{fig:St_alpha_grid}. 
    }
    \label{fig:St_beta_grid}
\end{figure}

In the limit of $\beta=0$, we have $\chi \rightarrow \atantwo [ (-1)/ (-2 \St)  ]$ as long as $\alpha$ is non-zero. 
When $\St \ll 1$, $\chi \rightarrow -90^{\circ}$, which would produce polarization parallel to the azimuthal direction. 
Fig.~\ref{fig:St_beta_grid}d shows $\chi$ in the limit of $\beta=0$. 
The reason is the following. 
With $\beta=0$, the azimuthal velocity of the gas is in Keplerian orbit, i.e., $v_{g,\Phi}=v_{K}$ (Fig.~\ref{fig:St_beta_grid}a). 
As the grains accrete with the gas from outer radii, they become super-Keplerian and feel a headwind towards $\chi=-90^{\circ}$ (Fig.~\ref{fig:St_beta_grid}c). In the case of large grains, $\St \gg 1$, they reach purely Keplerian orbits and only feel the flow from the radial direction due to the accreting gas. 
Although the flow speed is smaller by 1 order of magnitude compared to Fig.~\ref{fig:St_alpha_grid}h, it only lengthens the oscillation time $t_{o}$ and does not alter the damping time $t_{d}$. 
This limiting case highlights the fact that the direction of the gas flow can be complicated, especially in highly structured gas disks, and requires further investigation. 

\subsection{Alignment timescale} \label{sec:disk_alignment_timescale}

We make a simple prescription of a disk applying the velocity field formulated above. We assume a stellar mass of $M_{*} = 0.5$~$M_{\odot}$. The surface density is a power-law truncated at an inner radius $R_{a}$ and an outer radius $R_{b}$:
\begin{align} \label{eq:surface_density_profile}
    \Sigma(R) = \Sigma_{1\text{au}} \bigg( \frac{R}{ 1 \text{au} } \bigg)^{-p}
\end{align}
where $\Sigma_{1\text{au}}$ is the surface density at 1~au. We adopt $R_{a}=0.1$~au, $R_{b}=100$~au, and $p=1$. 
We assume a total disk mass that is $5\%$ of the stellar mass, $M_{\text{disk}} = 0.05$~$M_{*}$, which gives $\Sigma_{1\text{au}}\sim 350$~g cm$^{-2}$.  

The temperature of the disk follows
\begin{align} \label{eq:temperature_profile}
    T(R) = 200 \bigg( \frac{R}{\text{au}} \bigg)^{-0.5}. 
\end{align}

The pressure scale height is $H \equiv c_{s} / \Omega_{K}$. The gas density in the midplane is 
\begin{align}
    \rho_{g}(R) \equiv \frac{ \Sigma }{ \sqrt{2\pi} H } 
\end{align}
With this setup, the pressure gradient is $\beta=-2.75$. We adopt $\alpha=0.01$. 
For the prolate grain, we adopt $c=1$~mm, $\bar{a}=0.9$, and $g=0.01$.

\begin{figure}
    \centering
    \includegraphics[width=\columnwidth]{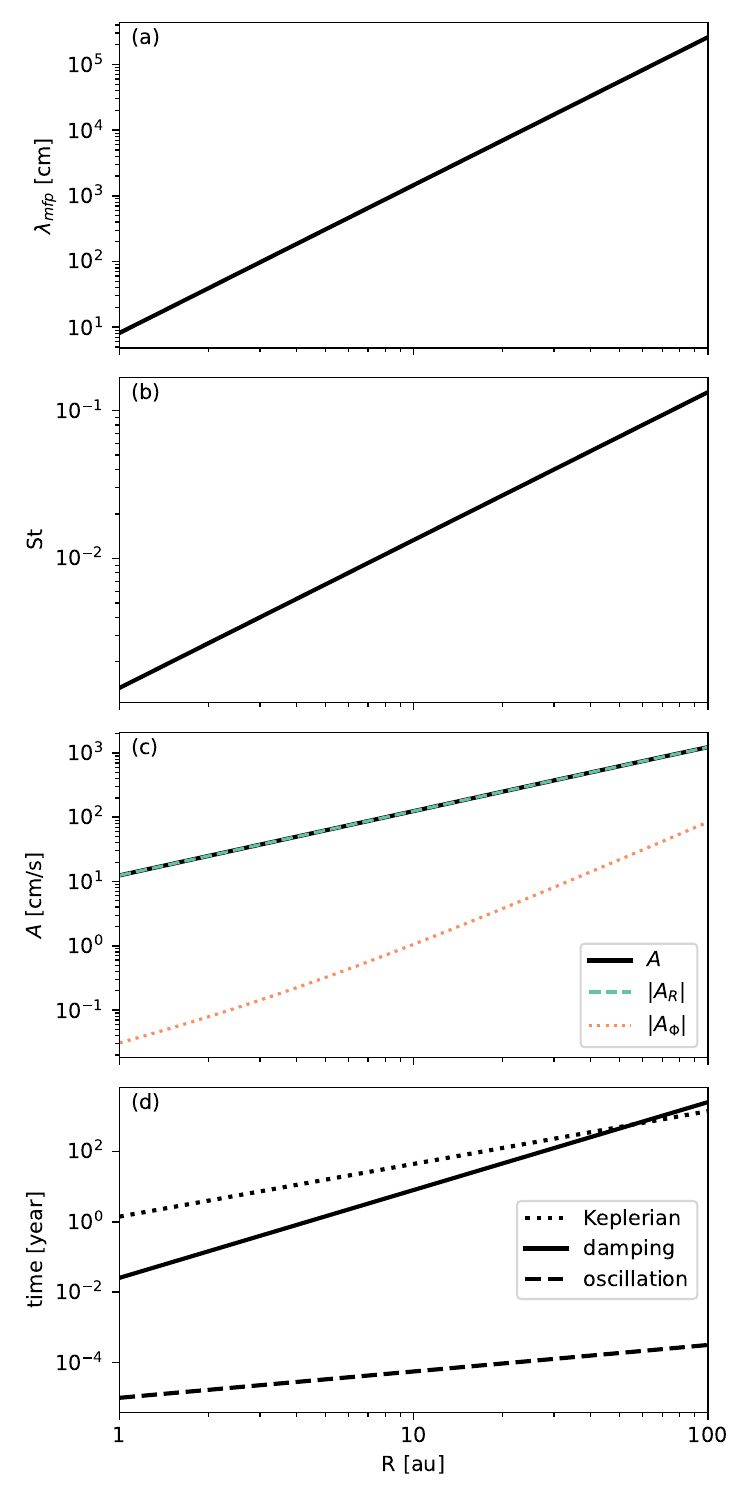}
    \caption{
        The resulting power-law profiles of a disk. 
        Panel a: the mean free path of the gas.
        Panel b: the $\St$. 
        Panel c: the dashed green and orange lines are the absolute values of 
        radial and azimuthal components of the gas flow, respectively. The black line is the magnitude of the gas flow $A$. 
        Panel d: the solid line is the damping time $t_{d}$, the dashed line is the oscillation time $t_{o}$, and the dotted line is the Keplerian orbital time $t_{K}$. 
    }
    \label{fig:time_profile_spheroid}
\end{figure}

For reference, we estimate the mean free path of the gas by 
\begin{align}
    \lambda_{\text{mfp}} = \frac{1}{ n_{g} \sigma_{\rm H_{2}} }
\end{align}
where $\sigma_{\rm H_{2}} = 2 \times 10^{-15}$~cm$^{2}$ \citep{Birnstiel2010A&A...513A..79B}. Fig.~\ref{fig:time_profile_spheroid}a shows $\lambda_{\text{mfp}}$ as a function of radius in the disk. For our adopted disk parameters, the $\lambda_{\text{mfp}}$ is $\sim 8$~cm at $R=1$~au to $\sim 2\times10^{5}$~cm at $=100$~au. Thus, the assumed millimeter grain should be well within the Epstein regime, especially in the lower-density regions like in the outer regions of a disk.

Fig.~\ref{fig:time_profile_spheroid}d shows the timescales as a function of radius in the disk. We find that the damping time $t_{d}$ is around 4~days ($\sim 10^{-2}$~year) at $R=1$~au, and reaches a thousand years at $R=100$~au. Also, $t_{d}$ is mostly smaller than the orbital time $t_{K}$ at all radii up to 70~au, meaning grain oscillations are mostly damped within a fraction of an orbit. The increase in $t_{d}$ with increasing radius is due to the decrease in both the midplane density and temperature (Eq.~\ref{eq:spheroid_t_d}). 
Given that $t_{d}$ is comparable to or less than $t_{K}$, we reason that the alignment mechanism can operate in disks even for a grain with a small offset between its center of mass and its geometric center of $g=0.01$. 

The oscillation time $t_{o}$ is around a few minutes ($\sim 10^{-5}$~year) at $R=1$~au and increases to an hour ($\sim 10^{-4}$~year) at $R=100$~au. 
Different from $t_{d}$, the oscillation time $t_{o}$ does not increase strongly with radius, because it is only proportional inversely to the square root of $\rho_{g}$ and in addition, benefits from an increased level of gas flow at larger radii. 
From the adopted parameters, the $\St$ is $\sim 10^{-3}$ at 1~au and increases with increasing radius to $\sim 0.1$ because the midplane density decreases while the grain size is fixed (Fig.~\ref{fig:time_profile_spheroid}b). 
The increasing $\St$ allows the grain to better decouple and experience stronger flow speed $A$ (Fig.~\ref{fig:time_profile_spheroid}c). 

As discussed previously for the disk and grain parameters chosen in this illustrative example, the direction of the flow is predominantly in the radial direction in the simple power-law prescription of a disk (Fig.~\ref{fig:time_profile_spheroid}c). More comprehensive calculations of the gas and dust velocity in structured disks are necessary to address the question of the alignment direction.

\section{Discussion} \label{sec:discussion}

\subsection{Comparison to other mechanical alignment mechanisms} \label{sec:comparison_to_other_alignment}

The badminton birdie-like alignment mechanism utilizes the grains' interactions with the ambient gas to achieve alignment, which makes it a type of mechanical (or aerodynamic) alignment. A popular type of mechanical alignment is alignment due to helicity, which is the ability to spin up a grain in the presence of gas flow \citep{Lazarian2007ApJ...669L..77L}. We showed that even without helicity, a grain can be aligned along the gas flow with the birdie-like alignment mechanism. These are two completely different effects. Our general torque equation, Eq.~\ref{eq:torque_three_terms_with_tensor}, likely captures the helicity effect through the diagonal terms of $\matr{M}$. While those terms are 0 for spheroids (Eq.~\ref{eq:spheroid_M}), if they are non-zero, the grains will obtain a torque in the same direction of $\vect{A}$ in grain body frame which corresponds to the definition of helicity. We leave a detailed exploration of the case of non-zero helicity to a future study.

Eq.~\ref{eq:torque_three_terms_with_tensor} also expects a driving torque from specular reflection if $\vect{K}$ is non-zero. 
Recall that for spheroids, $\vect{K} = \vect{0}$. 
However, if $\vect{K}$ exists (depending on the grain geometry), then gas pressure can generate a torque in the grain frame and spin up the grain. 
The torque does not require a gas flow, making it different from the flow-induced restoring torque from the birdie-like alignment process. 
This is similar to the systematic torques proposed by \cite{Purcell1979ApJ...231..404P}, but they differ in physical origin. 
Interestingly, the driving torque goes as $\propto c_{s}^{2}$ which is much larger than the restoring torque (related to $\matr{M}$) which is $\propto \vth A$. 
$\vect{K}$ would have to be very small to not be effective. 
If $\vect{K}$ can increase the rotational energy of the grain beyond the escape energy ($2 P_{s}$; see Sec.~\ref{sec:doublesphere} and \ref{sec:spheroid}), birdie-like alignment effect may be overwhelmed by the rapid spin. 
Nevertheless, since $\vect{K}$ is determined entirely from the surface features of a grain which we expect to be random in general, we speculate that even if $\vect{K}$ is the dominant component for certain grains, the alignment direction should also be random as an ensemble and produce a net zero polarization. 
We leave a more comprehensive exploration of the non-zero $\vect{K}$ case to a future study.

Another type of mechanical alignment is the Gold alignment mechanism \citep{Gold1952MNRAS.112..215G, Gold1952Natur.169..322G}. 
We can describe the polarization from the Gold alignment mechanism through two main parts: the first being the actual Gold alignment part and the second being the polarization from the projection of grains once they are Gold-aligned. 
The first part utilizes prolate grains under the presence of a flow of gas particles. Since the gas particles predominantly impact the grains along the long axis (the other two axes are negligible), the angular momentum is entirely parallel to the short axis, and the angular momentum is mostly confined to a plane perpendicular to the gas flow. 
As a collection, the angular momentum directions are random around the direction of the flow and, for any instance, the directions of the long axes of the grains are random around their own angular momentum. 
The second part is the polarization that is a natural result of projection of the collection of prolate grains with this configuration. 
Polarization mostly comes from the grains whose angular momenta are perpendicular to the observer as long as the flow is not directed to the observer.

Our formulation of Eq.~\ref{eq:torque_three_terms_with_tensor} does not immediately produce the results of Gold alignment. The grain structure from \cite{Gold1952MNRAS.112..215G} corresponds to a prolate which has $\vect{K}=\vect{0}$ and also the center of mass is the same as the geometric center, making $\matr{M}=\matr{0}$. 
As a result, the torque from $\vect{A}$ is always $\vect{0}$ regardless of how large $\vect{A}$ is. 
Therefore, we would expect that spinning grains damped by the gas should eventually stop at an orientation that is entirely determined by the initial conditions (see Eq.~\ref{eq:solution_P_eq_0}). 
One may notice that from Eq.~\ref{eq:solution_P_eq_0} the spin axis will be around $\vect{\hat{e}}_{2}$, which is defined to be perpendicular to $\vect{A}$, and we may obtain polarization from the Gold-projection effect (second part). However, the polarization is an artificial result because we have assumed that grains can only oscillate around $\vect{\hat{e}}_{2}$. In 3D, with $\matr{M}=\matr{0}$, the orientation of angular momenta and the resulting final orientation should be entirely random with respect to $\vect{A}$.

One possible reconciliation is that we have assumed that the speed of each patch of the surface is small compared to the sound speed of gas (Appendix~\ref{sec:force_per_unit_surface_derivation}). 
Numerous studies have found that the Gold alignment requires supersonic drifts \citep{Gold1952MNRAS.112..215G, Purcell1969Phy....41..100P, Lazarian1994MNRAS.268..713L}. 
Thus, it appears consistent with Gold alignment that Eq.~\ref{eq:torque_three_terms_with_tensor} should not produce a torque for any $\vect{A}$ without the offset between the center of mass and geometric center. 
We suspect that Gold alignment corresponds to the supersonic version of Eq.~\ref{eq:torque_three_terms_with_tensor}, but leave the verification to a future investigation.

\subsection{Does birdie-like alignment work in the ISM or protostellar envelopes?} \label{sec:birdie_like_alignment_in_ISM}

We estimate the damping and oscillation timescales for a grain in ISM conditions. 
Using Eq.~\ref{eq:spheroid_t_d}, the damping time is 
\begin{align} \label{eq:t_ds_ism}
    t_{d,s} \sim 3.5 \times 10^{7} \text{years} 
    \bigg( \frac{\rho_{s}}{ 3 \text{g cm}^{-3}} \bigg)
    \bigg( \frac{c}{0.1 \mu\text{m}} \bigg)
    \bigg( \frac{ 20 \text{cm}^{-3} }{ n_{g} } \bigg)
    \bigg( \frac{0.3 \text{km /s}}{\vth} \bigg)
\end{align}
when adopting $\bar{a}=0.9$ and $\bar{g}=0.01$; again, $c$ is the grain size, which we normalize by the classic size of 0.1~$\mu$m for the ISM.
Using Eq.~\ref{eq:spheroid_t_o}, the oscillation time is 
\begin{align} \nonumber
    t_{o,s} \sim 110 \text{hours} 
        \sqrt{ \frac{ \rho_{s} }{3 \text{g cm}^{-3} } } 
        \bigg( \frac{ c }{ 0.1 \mu\text{m} } \bigg)
        \sqrt{\frac{ 20 \text{cm}^{-3} }{ n_{g} } }
        \sqrt{ \frac{ 0.3 \text{km/s}  }{ \vth } }
        \sqrt{ \frac{ 1 \text{cm/s} }{ A } }
\end{align}
also using $\bar{a}=0.9$ and $\bar{g}=0.01$. 
We can easily see that the low-density environment means the grains will oscillate rapidly, while it takes a long time to damp out the oscillations.

For the ISM, however, there can be other processes that can systematically spin up the grain to rates comparable to thermal rotation, like radiative torques \citep{Draine1996ApJ...470..551D, Hoang2009ApJ...697.1316H}. We can parameterize the rotational energy from other spin-up processes as $f k T$ where $f$ is a multiplication factor and $kT$ is the thermal energy. If the grains are in thermal equilibrium, we would expect $f=3/2$.
We provide an estimation of how much flow is necessary to trap the grain into oscillation using $2 P_{s} > f k T$. The relation gives a threshold flow speed $A_{t,s}$: 
\begin{align} \label{eq:A_ts}
    A > A_{t,s} &\equiv \frac{f \vth }{n_{g} c^{3}} \Breve{A}_{t,s} \text{ ,} \nonumber \\
    \Breve{A}_{t,s} &\equiv \frac{1}{16 \bar{a} \bar{g} E[1 - x^{2}] }
\end{align}
where $\Breve{A}_{t,s}$ is a dimensionless factor and the physical combination is the characteristic threshold drift speed. 
For $\bar{a}=0.9$ and $\bar{g}=0.01$, we get $\Breve{A}_{t,d}\sim 5$. 

Adopting the same parameters used for Eq.~\ref{eq:t_ds_ism}, we get an impractical $A_{t,d} = 8 f \times 10^{18}$~cm/s. 
In contrast, $A_{t,d} \sim 0.5$~cm/s for a disk using the conditions from Eq.~\ref{eq:double_sphere_stopping_time} with the same $\bar{a}$ and $\bar{g}$. 
The main issue is that the threshold flow speed is very sensitive to the grain size $c$, with $A_{t,s} \propto c^{-3}$, meaning it is much harder to trap small grains. 
The secondary issue is that the density is much lower in the ISM than in protoplanetary disks. 
Thus, the birdie-like mechanism is unlikely to compete with other mechanisms that can provide thermal rotational energy to grains in the ISM. 

We next assess the dense cores of molecular clouds, since dust polarization is routinely measured in dense cores and protostellar envelopes \citep[e.g.][]{Stephens2013ApJ...769L..15S, Maury2018MNRAS.477.2760M, Galametz2018A&A...616A.139G, LeGouellec2019ApJ...885..106L, Yen2020ApJ...893...54Y, Pattle2021ApJ...907...88P, Cox2022ApJ...932...34C, LeGouellec2023A&A...671A.167L, Lin2024ApJ...961..117L, Huang2024ApJ...963L..31H}. 
We find that the damping time is 
\begin{align} \label{eq:t_ds_core}
    t_{d,s} \sim 7 \times 10^{4} \text{years} 
    \bigg( \frac{\rho_{s}}{ 3 \text{g cm}^{-3}} \bigg)
    \bigg( \frac{c}{1 \mu\text{m}} \bigg)
    \bigg( \frac{ 10^{5} \text{cm}^{-3} }{ n_{g} } \bigg)
    \bigg( \frac{0.3 \text{km/s}}{\vth} \bigg)
\end{align}
where we adopted $c=1$~$\mu$m to account for some grain growth from classic ISM values of 0.1~$\mu$m \citep[e.g.][]{Pagani2010Sci...329.1622P} and we assumed $\bar{a}=0.9$ and $\bar{g}=0.01$. 
The timescale appears less than or comparable to the lifetimes of the Class~0/I sources embedded in envelopes \citep{Evans2009ApJS..181..321E, Williams2011ARA&A..49...67W}. 
Applying the same conditions, the oscillation time is
\begin{align} \nonumber
    t_{o,s} \sim 16 \text{hours} 
        \sqrt{ \frac{ \rho_{s} }{3 \text{g cm}^{-3} } } 
        \bigg( \frac{ c }{ 1 \mu\text{m} } \bigg)
        \sqrt{\frac{ 10^{5} \text{cm}^{-3} }{ n_{g} } }
        \sqrt{ \frac{ 0.3 \text{km/s}  }{ \vth } }
        \sqrt{ \frac{ 1 \text{cm/s} }{ A } }. 
\end{align}
The value for $A$ is unclear, but it can be much less than the adopted value if the small grains are well coupled to the gas.
Following Eq.~\ref{eq:A_ts}, we find that $A_{t,s} \sim 1.5 f \times 10^{12}$~cm/s. 
The large threshold suggests that birdie-like alignment is unlikely to trap grains in oscillation in dense cores of molecular clouds or protostellar envelopes if other processes can drive thermal rotation unless the grains in such regions have already grown much larger than 1~$\mu$m (as indicated by multi-frequency dust continuum observations in some cases, e.g., \citealt{Kwon2009ApJ...696..841K})


\subsection{Connection to observations} \label{sec:connection_to_observations}

The drag force from gas plays an important role in the disk dynamics of dust and is a key ingredient to explain dust rings due to pressure bumps \citep[e.g.][]{Pinilla2012A&A...538A.114P, Dullemond2018ApJ...869L..46D}, the streaming instability \citep[e.g.][]{Youdin2005ApJ...620..459Y, Squire2020MNRAS.498.1239S}, or pebble accretion \citep{Ormel2017ASSL..445..197O}. 
The birdie-like alignment mechanism makes continuum polarization observations direct evidence of gas drag and offers a possibility to probe parts of the dust kinematics. 
While the Doppler shifts of molecular line emission can trace the gas kinematics, empirical constraints on the dust kinematics are harder to come by. 
Our work adds to the idea that continuum polarization can measure the relative velocities between the gas and dust \citep{Kataoka2019ApJ...874L...6K, Mori2019ApJ...883...16M, Tang2023ApJ...947L...5T}, but we offer a new physical justification that grains can be aligned along the long axis even with highly subsonic drift. 

Several disks show polarization that is elliptical in the polarization orientation with HL~Tau as the most prominent example seen from the multiwavelength polarization \citep{Stephens2017ApJ...851...55S, Lin2024MNRAS.528..843L} and the high angular resolution polarization \citep{Stephens2023Natur.623..705S}. If polarization is indeed due to birdie-aligned grains, the $\vect{A}$-field should be in the azimuthal direction with a $180^{\circ}$ degeneracy. 

Intriguingly, there are two disks with a spiral pattern, namely AS~209 \citep{Mori2019ApJ...883...16M} and GG~Tau \citep{Tang2023ApJ...947L...5T} that are relevant to our discussion. The polarization patterns for both are nearly azimuthal, but with a slight deviation by $4.5^{\circ}$ and $7.1^{\circ}$, respectively. The direction of deviation for both follows an inward spiral. For both sources, the conventional alignment of grains with helicity cannot easily explain the inward spiral and they favor a scenario where prolate grains drift inwards and in the azimuthal direction, which is better explained by the Gold mechanism \citep{Mori2019ApJ...883...16M, Tang2023ApJ...947L...5T}. The immediate benefit from the birdie-like alignment is that while the Gold mechanism requires supersonic drift, the birdie-like alignment only requires subsonic drift and reproduces the same orientation that the Gold mechanism offers. 
Whether or not the birdie-like alignment can produce the inward spiral depends on the proper treatment of the gas and dust velocity in a disk, or around a ring for these two sources in particular, which is beyond the scope of this paper.



\subsection{Caveats and Future Developments} \label{sec:caveats_and_future}

There are several assumptions made in this exploratory work. 
First, we assume that angular momentum is predominantly around $\vect{\hat{e}}_{2}$ (the direction of the gas flow-induced torque) when, generally, we might expect angular momentum around the other two axes as well. 
In the general case, the rotating motion around $e_2$ (and $b_2$ considered in this paper) will become precession in 3D. 
For example, if the grain contains significant angular momentum around $\vect{\hat{b}}_{3}$, $\Gamma_{2}$ will induce precession around $\vect{\hat{e}}_{3}$. 
However, we can expect that the damping torque should inevitably diminish the spin and precession at timescales of order $t_{d}$ (which is rather short in the high-density disk environment where the birdie alignment is most applicable) and reach alignment akin to what is explored in this work. 
Note that a spheroid will not exhibit drag when spinning around its axis of asymmetry, but it should exist for realistic, irregular grains. 
Another side effect of the assumption is that grains may artificially exhibit polarization simply from the Gold-projection effect in the unlikely limiting case of $\matr{M}=\matr{0}$ as described in Sec.~\ref{sec:comparison_to_other_alignment}. 

Another assumption is that the level of $\vect{A}$ is fixed as the grain oscillates. However, the geometric cross-section in the direction of $\vect{A}$ varies as the grain oscillates. 
Since we know the force on each patch of the surface, it is a simple exercise to derive the total force on the axisymmetric spheroid using Eq.~\ref{eq:force_per_unit_area}, Eq.~\ref{eq:solid_body_velocity_difference}, and Eq.~\ref{eq:spheroid_surface_radius_vector}:
\begin{align} \label{eq:force_linear_motion}
    \vect{F} = \oint_{S} d \vect{F} = - \rho_{g} \vth \oint_{S} \matr{N} d \sigma (\vect{\omega} \times \vect{g} - \vect{A}). 
\end{align}
For a spheroid, we know $\oint_{S} \matr{N} d\sigma$ from Eq.~\ref{eq:oint_N_dsigma}. 
For the case of a sphere, $\vect{g}=\vect{0}$ and $\bar{a}=1$, we recover the Epstein drag of a sphere with specular reflection (Eq.~\ref{eq:epstein_drag}; note that $\vect{u} = - \vect{A}$ for that equation). 
We can identify in Eq.~\ref{eq:force_linear_motion} that $\vect{\omega} \times \vect{g} - \vect{A}$ is the velocity of the center of the spheroid relative to the gas as it oscillates around the center of mass. 
The relation means that the force that determines the translational motion of the grain can also depend on the rotational motion and a fully self-consistent study of the velocity of the grain in a disk requires incorporating the equation of rotational motion. Nevertheless, since we are particularly interested in the scenario when $\vect{\omega}$ is small (near alignment) and we found that $\vect{g}$ only needs to be $\sim 1\%$ of the length scale for alignment, we do not expect the flow to vary much with oscillation. 

The last assumption of importance is the grain structure. 
We utilized two cases of an axisymmetric spheroid: the strictly axisymmetric case where the offset between the geometric and mass centers is along the axis of symmetry and the quasi-axisymmetric case where the offset can be anywhere in the spheroid. 
The assumption allowed us to study alignment analytically. 
To explain the disk polarization observations, the center of mass should be shifted along the long axis of the prolate or oblate instead of the short axis even though birdie-like alignment does permit prolates or oblates to be aligned with $\vect{A}$ along the short axes. 
In other words, observations suggest that the predominant grain structure favors the offset vector $\vect{g}$ to be along the long axes. 
It is unclear how grain growth can lead to this structure, but we can view this as a result that the grain growth mechanism should strive to explain. 
Our formulation of Eq.~\ref{eq:torque_three_terms_with_tensor} allows one to precalculate the dynamical properties of the grain ($\vect{K}$, $\matr{L}$, and $\matr{M}$) independent of environmental properties or dynamical state. Eq.~\ref{eq:torque_three_terms_with_tensor} should now allow a direct connection between grain structure and its alignment permitting studies of several geometries quickly. 
This is a very attractive possibility as we can study the alignment of a population of grains each with a unique geometry, allowing the prediction of the degree of alignment and thus formulate testable predictions of polarization for various grain growth mechanisms. 
We leave a more detailed exploration of these assumptions to future studies. 

\section{Conclusions} \label{sec:conclusion}

Disk scale polarization has shown evidence of effectively prolate grains aligned azimuthally. 
However, the alignment mechanism is unclear. 
In this paper, we demonstrate the possibility that grains can be aligned by the gas flow when the center of mass is offset from its geometric center, as in the case of a badminton birdie. 
Our main results are summarized as follows: 
\begin{enumerate}
    \item To build our physical intuition, we first utilize a simple model of a grain that is composed of two spheres connected by a rigid, massless rod. We show that the grain behaves as a damped oscillator under the presence of a systemic flow of gas, $\vect{A}$, if the two spheres of the grain are not identical. 
    The oscillation is due to restoring torques generated by $\vect{A}$ and the asymmetry between the spheres, i.e., the ``flow-induced restoring torques'' which resists the angular displacement from the direction of alignment.
    The oscillation can be described by a potential well near the direction of alignment. As the gas damps the grain oscillation, the grain reaches alignment such that the geometric center follows the direction of the flow with respect to its center of mass. We derive the damping time and oscillation time and show that the oscillation time is much shorter than the damping time for typical circumstellar disk conditions (i.e., underdamped oscillations). The damping time scale is set by the dust stopping time and the dust properties, particularly the degrees of grain elongation and asymmetry.
    \item We derive a formula for the torque of a smooth body in the Epstein regime by considering subsonic relative motion and specular reflection of gas on each patch of the surface (Eq.~\ref{eq:torque_three_terms_with_tensor}). 
    We capture the surface properties of the grains relevant for the grain dynamics by $\vect{K}$, $\matr{L}$, and $\matr{M}$ which characterizes the driving torque from gas pressure, the damping torque as the grain spins, and the torque that arises from the existence of $\vect{A}$, respectively. These quantities allow us to quantify the grain rotational motion without predetermining environmental properties, like the gas density or temperature, and the dynamic state, like the orientation or angular velocity. 
    \item We apply the torque equation to an axisymmetric spheroid (prolate and oblate) and obtain the damped oscillator equation of motion along with analytical solutions to the damping time and oscillation time. By considering a grain whose center of mass is shifted along the axis of symmetry, which is characterized by $\vect{g}$ (the offset vector from the center of mass to the center of the spheroid), we find that the final alignment direction is such that $\vect{g}$ is parallel to $\vect{A}$. In other words, for a prolate, the long axis is along the gas flow, or for an oblate, the short axis is along the gas flow. 
    \item We further consider a grain whose center of mass can be anywhere within the spheroidal surface and also derive analytical solutions to the damping time and oscillation time. We show that the final alignment direction depends on both $\vect{g}$ and the shape of the spheroid. In particular, if the center of mass of an oblate is shifted along the long axis, then the oblate will be aligned along the long axis. Since disk polarization observations require effectively prolate grains, the birdie-like alignment mechanism requires the offset vector $\vect{g}$ to be along the long axis for both prolates and oblates. 
    \item We implement a simple disk model and calculate the gas and dust velocity fields. Using the power-law disk model with a pressure gradient, we expect the gas flow to be in the radial direction for small Stokes number, $\St$, and in the azimuthal direction for large $\St$. However, in the center of pressure bumps or gaps, where the pressure gradient can be small, the gas flow can be in the azimuthal direction for small $\St$ and in the radial direction for large $\St$. Using just the power-law disk model to study the radial dependence of the mechanism, we show that the damping time is mostly less than the Keplerian time $t_{K}$, but it can be comparable to or greater than $t_{K}$ at large radii near the disk outer edge. Typically, it takes a fraction of an orbit or only several orbits to align a grain in a disk. The very short alignment timescale compared to the lifetime of the disk makes birdie-like alignment a very promising mechanism to explain the inferred toroidally aligned, effectively prolate grains from several disk polarization observations and deserves further development. 
\end{enumerate}

\section*{Acknowledgements}
We thank the anonymous reviewer for the constructive comments that improved the paper.
Z.-Y.D.L. acknowledges support from NASA 80NSSC18K1095, the Jefferson Scholars Foundation, the NRAO ALMA Student Observing Support (SOS) SOSPA8-003, the Achievements Rewards for College Scientists (ARCS) Foundation Washington Chapter, the Virginia Space Grant Consortium (VSGC), and UVA research computing (RIVANNA).
ZYL is supported in part by NASA 80NSSC20K0533 and NSF AST-2307199.
L.W.L. acknowledges support from NSF AST-1910364 and NSF AST-2307844.
MFL acknowledges support from the European Research Executive Agency HORIZON-MSCA-2021-SE-01 Research and Innovation programme under the Marie Skłodowska-Curie grant agreement number 101086388 (LACEGAL). MFL is also grateful for the warmth of the Star Formation group of the Institut de Ci\`encies de l'Espai and the Universidad de Barcelona during his 2024 research stay.

\section*{Data Availability}

Data underlying this article is available from the corresponding author upon request.



\bibliographystyle{mnras}
\bibliography{main} 




\appendix

\section{Additional Example of a Phase Portrait} \label{sec:better_damped_phase_portrait}

\begin{figure*}
    \centering
    \includegraphics[width=0.9\textwidth]{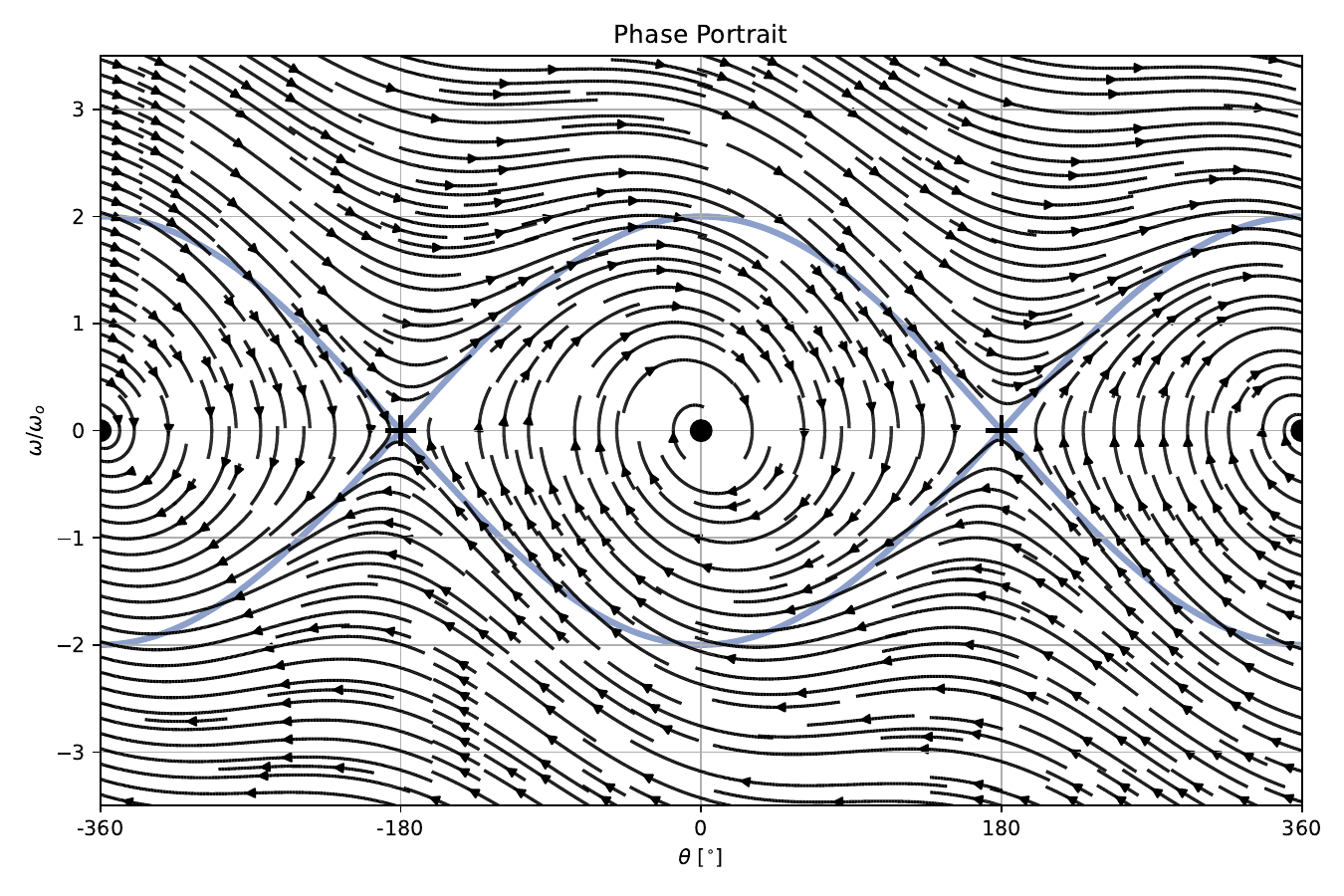}
    \caption{
        The phase portrait plotted in the same way as Fig.~\ref{fig:doublesphere_phase_portrait}. In this example, the grain is more damped. 
    }
    \label{fig:doublesphere_phase_portrait_2}
\end{figure*}

Fig.~\ref{fig:doublesphere_phase_portrait} from Sec.~\ref{sec:doublesphere} presented a phase portrait showing the damped oscillation behavior of a double-sphere model of a grain. The typical disk conditions mean grains undergo highly underdamped oscillations. To illustrate how a better-damped phase portrait appears, we adopt $n_{g}=10^{12}$~cm$^{-3}$, $\vth=0.2$~km/s, $A=5\times 10^{-5}$~cm/s, $l=1$~mm, $\epsilon=1$, and $\kappa=1.01$ and the results are shown in Fig.~\ref{fig:doublesphere_phase_portrait_2}. With a much higher gas density and smaller velocity difference, the trajectories now clearly spiral toward the attractors (final alignment points), even starting from initially unbound states considered in this plot. For reference, using Eq.~\ref{eq:damping_ratio}, the damping ratio $\zeta_{d} \sim 0.14$ in this illustrative case. 

\section{Derivation of the drag force on a surface element} \label{sec:force_per_unit_surface_derivation}

In this appendix, we assume rarefied, subsonic gas flow (i.e., Epstein regime) and obtain the drag force on a patch of surface. The validity of the Epstein regime in a disk is discussed in Sec.~\ref{sec:disk_alignment_timescale}. We will use a distribution of velocities of the gas for a moving observer and derive the transfer of momentum from the gas onto a patch of surface after integrating over the velocity distribution.
The derivation is based on \cite{Epstein1924PhRv...23..710E}. We will redefine the symbols utilized in this appendix for clarity and the same symbols in the main text may not correspond to the definition here. 

We begin by considering the Maxwell velocity distribution for the gas which gives the number of molecules having velocity components in the cartesian coordinates between $\xi_{0}$, $\eta_{0}$, $\zeta_{0}$ and $\xi_{0} + d \xi_{0}$, $\eta_{0} + d \eta_{0}$, $\zeta_{0} + d \zeta_{0}$ through:
\begin{align}
    N_{\xi_{0}, \eta_{0}, \zeta_{0}} d \xi_{0} d \eta_{0} d \zeta_{0}
    = N \bigg( \frac{h}{\pi} \bigg)^{3/2} e ^{-h(\xi_{0}^{2} + \eta_{0}^{2} + \zeta_{0}^{2})} d \xi_{0} d \eta_{0} d \zeta_{0}
\end{align}
where $N$ is the number of molecules per unit volume. $h$ is defined by 
\begin{align}
    h \equiv \frac{m}{2 k T} \nonumber
\end{align}
where $m$ is the mass of each molecule, $k$ is the Boltzmann constant, and $T$ is the temperature. 

Consider a traveling observer with a cartesian coordinate system with axes $x$, $y$, and $z$. 
The observer travels relative to the rest frame of the gas with a speed of $V$ and components $\alpha V$, $\beta V$, and $\gamma V$ along the $x$, $y$, and $z$ axes, respectively. Here, $\alpha$, $\beta$, and $\gamma$ are direction cosines (not to be confused with $\alpha$ and $\beta$ from Sec.~\ref{sec:disk}).
Let $\xi$, $\eta$, and $\zeta$ be the velocity components of a molecule parallel to $x$, $y$, and $z$, respectively, seen by the observer. The transformation between the velocity in the observer frame to the velocity in the gas rest frame follows
\begin{align}
    \xi = \xi_{0} - \alpha V \text{ , } \nonumber \\
    \eta = \eta_{0} - \beta V \text{ , } \nonumber \\
    \zeta = \zeta_{0} - \gamma V \text{ . }
\end{align}
Since $V$ is constant, $d \xi = d \xi_{0}$, $d \eta = d \eta_{0}$, and $d \zeta = d \zeta_{0}$. Thus, the velocity distribution seen by the observer is 
\begin{align}
    &N_{\xi, \eta, \zeta} d \xi d \eta d \zeta = \nonumber \\
    &N \bigg( \frac{h}{\pi} \bigg)^{3/2} e ^{-h[(\xi+\alpha V)^{2} + (\eta+\beta V)^{2} + (\zeta+\gamma V)^{2}]}
    d \xi d \eta d \zeta
\end{align}
We approximate the equation by considering the distribution to the first power of $V$ which gives
\begin{align}
    N_{\xi, \eta, \zeta} 
        = N \bigg( \frac{h}{\pi} \bigg)^{\frac{3}{2}}
        \bigg[1 - 2 h V (\alpha \xi + \beta \eta + \gamma \zeta) \bigg] e^{-h (\xi^{2} + \eta^{2} + \zeta^{2})}
    \text{ .}
\end{align}

We now wish to calculate the number of impinging particles onto some surface element $d \sigma$. Let the $x$-axis be the direction that is normal to the surface element. Within some unit time $dt$, the impinging particles with velocity ($\xi$, $\eta$, $\zeta$) will be those that are enclosed within a cylinder with $d \sigma$ as the base and $-\xi d t$ as the height: 
\begin{align} \label{eq:n_impinging_particles}
    &n_{\xi, \eta, \zeta} d \xi d \eta d \zeta dt d \sigma = \\ &-N \bigg( \frac{h}{\pi} \bigg)^{\frac{3}{2}} 
        \bigg[\xi - 2 h V \xi (\alpha \xi + \beta \eta + \gamma \zeta) \bigg] e^{-h (\xi^{2} + \eta^{2} + \zeta^{2})}
        d \xi d \eta d \zeta dt d \sigma  \nonumber \text{ .}
\end{align}

Each imping particle carries a momentum of 
\begin{align}
    \vect{p}_{i} = m \begin{pmatrix}
        \xi \\
        \eta \\
        \zeta
    \end{pmatrix}
\end{align}
in the observer frame. Since we know the number of impinging particles onto the surface from Eq.~\ref{eq:n_impinging_particles}, we can obtain the momentum from the impacts of all velocities 
\begin{align}
    \vect{p}^{(i)}_{\sigma, t} d t d \sigma &= \int_{- \infty}^{0} d \xi \int_{- \infty}^{\infty} d \eta \int_{-\infty}^{\infty} d \zeta \vect{p}_{i} n_{\xi, \eta, \zeta} d t d \sigma \\
    &= - m N \Bigg[
        \begin{pmatrix}
            \frac{1}{4h} \\
            0 \\
            0
        \end{pmatrix}
        + \frac{V}{2 \sqrt{\pi h}}
        \begin{pmatrix}
            2 \alpha \\
            \beta \\
            \gamma
        \end{pmatrix}
    \Bigg] 
    d t d \sigma
\end{align}
where $\vect{p}^{(i)}_{\sigma, t}$ is the momentum per unit area and per unit time.

To calculate the momentum of the emerging gas, we assume specular reflection in which, after the impact, the velocity of the molecule normal to the surface flips by a negative sign. The momentum of the emerging molecule is thus $\vect{p}_{e} = m (-\xi, \eta, \zeta)$. Following the same exercise as above, 
\begin{align}
    \vect{p}^{(e)}_{\sigma, t} d t d \sigma &= \int_{- \infty}^{0} d \xi \int_{- \infty}^{\infty} d \eta \int_{-\infty}^{\infty} d \zeta \vect{p}_{e} n_{\xi, \eta, \zeta} d t d \sigma \\
    &= - m N \Bigg[
        \begin{pmatrix}
            - \frac{1}{4h} \\
            0 \\
            0
        \end{pmatrix}
        + \frac{V}{2 \sqrt{\pi h}}
        \begin{pmatrix}
            - 2 \alpha \\
            \beta \\
            \gamma
        \end{pmatrix}
    \Bigg]
    d t d \sigma.
\end{align}
We can obtain the force onto the surface by looking at the momentum that is deposited by gas within the unit time: 
\begin{align} \label{eq:force_per_unit_area_derivation}
    d \vect{F} &\equiv \frac{[\vect{p}^{(i)}_{\sigma, t} - \vect{p}^{(e)}_{\sigma, t} ] d t d \sigma }{dt} \nonumber \\
    &= - m N \bigg( \frac{1}{2h} + \frac{2 V \alpha }{\sqrt{\pi h}} \bigg) \vect{n} d \sigma
\end{align}
where $d \vect{F}$ is the force on $d \sigma$ and $\vect{n}$ is the unit direction along the $x$-axis, which was defined to be in the normal direction of the surface. One can quickly realize that the force is only in the normal direction because the momentum of the molecules along the $y$- or $z$-directions has not changed. Also, the direction of the force is anti-parallel to $\vect{n}$. 

We can re-express the equation by connecting it to a few physical quantities. $1/2h$ is simply the isothermal sound speed squared $c_{s}^{2} \equiv k T / m$. Also, $2 / \sqrt{\pi h}$ is the average speed $\vth \equiv \sqrt{8 k T / (\pi m)}$. 
The quantity $m N$ is the mass density $\rho_{g}$ as defined in the main text. Furthermore, since $V \alpha$ is the velocity component along the $x$-axis, we can express it as $V \alpha = \vect{V} \cdot \vect{n}$. When expressed in the dot product form and applied to $\vect{n} d \sigma$ for Eq.~\ref{eq:force_per_unit_area_derivation}, one can quickly see that $(\vect{V} \cdot \vect{n}) \vect{n}$ is simply a vector from $\vect{V}$ that is projected onto the normal direction. We can express this relation through a projection tensor $\matr{N} \equiv \vect{n} \vect{n}$ giving $(\vect{V} \cdot \vect{n}) \vect{n} = \matr{N}(\vect{V})$. 
Implementing these new relations, we have
\begin{align}
    d \vect{F} = - \rho_{g} \bigg[ c_{s}^{2} \vect{n} + \vth \matr{N}(\vect{V}) \bigg] d \sigma
\end{align}
which matches Eq.~\ref{eq:force_per_unit_area}. 
Note that the first term is simply the pressure which does not depend on $\vect{V}$. 
The second term controls the effect of the motion of the surface.
When $\vect{V}$ is in the same direction as $\vect{n}$, the side of the surface that experiences reflection feels an extra force in the direction opposite to $\vect{n}$. When $\vect{V}$ is in the opposite direction of $\vect{n}$, the surface feels a reduced force. 

\section{Analytical Solutions to the $E$-Integrals} \label{sec:E_solutions}

In the main text, Eq.~\ref{eq:integral_E} defines an integration form that frequently appears in the equations of motion in Sec.~\ref{sec:spheroid}. The actual quantities that appeared are $E[x^{2}]$, $E[1 - x^{2}]$, and $E[x^{2} - x^{4}]$. Note that $E[f(x) + g(x)] = E[f(x)] + E[g(x)]$ where $f$ and $g$ are some functions of $x$. The property means we only need to solve for $E[1]$, $E[x^{2}]$, and $E[x^{4}]$. Here we provide the analytical solutions. 

For convenience, we can define the elongation factor for prolates (when $\bar{a} < 1$) as 
\begin{align}
    e \equiv \sqrt{1 - \bar{a}^{2}}
\end{align}
and the elongation factor for oblates (when $\bar{a}>1$) as 
\begin{align}
    \varepsilon \equiv \sqrt{\bar{a}^{2} - 1}
\end{align}
(not to be confused with $\epsilon$ in Sec.~\ref{sec:doublesphere}). Both are positive quantities. We find that 
\begin{equation} \label{eq:E_x0}
    E[1] = \begin{cases}
        \frac{2 \arcsin{(e)}}{ e } & \text{ , when } \bar{a} < 1 \\
        2 & \text{ , when } \bar{a} = 1 \\
        \frac{2 \arcsinh{(\varepsilon)}}{ \varepsilon } & \text{ , when } \bar{a} > 1 
    \end{cases}
\end{equation}
where $\arcsinh$ is the inverse hyperbolic sine. Furthermore,
\begin{equation} \label{eq:E_x2}
    E[x^{2}] = 
    \begin{cases}
        - \frac{ \sqrt{1 - e^{2}} }{ e^{2} } + \frac{ \arcsin{(e)} }{ e^{3} } & \text{ , when } \bar{a} < 1 \\
        \frac{2}{3 } & \text{ , when } \bar{a} = 1 \\
        \frac{ \sqrt{1 + \varepsilon^{2}} }{ \varepsilon^{2} } - \frac{ \arcsinh{(\varepsilon)} }{ \varepsilon^{3} } & \text{ , when } \bar{a} > 1 . \\
    \end{cases}
\end{equation}
Lastly, we can obtain
\begin{equation} \label{eq:E_x4}
    E[x^{4}] = 
    \begin{cases}
        - \frac{  (2 e^{2} + 3) \sqrt{1 - e^{2}}}{ 4 e^{4} } + \frac{ 3 \arcsin{(e)} }{ 4 e^{5} } & \text{ , when } \bar{a} < 1 \\
        \frac{2}{5 } & \text{ , when } \bar{a} = 1 \\
        \frac{ (2 \varepsilon^{2} - 3) \sqrt{1 + \varepsilon^{2}} }{ 4 \varepsilon^{4} } + \frac{ 3 \arcsinh{(\varepsilon)} }{ 4 \varepsilon^{5} } & \text{ , when } \bar{a} > 1 .\\
    \end{cases}
\end{equation}
With Eq.~\ref{eq:E_x0}, \ref{eq:E_x2}, and \ref{eq:E_x4}, we can get $E[1-x^{2}] = E[1] - E[x^{2}]$ and $E[x^{2} - x^{4}] = E[x^{2}] - E[x^{4}]$.


\bsp	
\label{lastpage}
\end{document}